\documentclass[a4paper,11pt]{article}

\usepackage{jheppub} 
\usepackage[T1]{fontenc} 
\usepackage[italian,english]{babel}
\usepackage{hyperref}
\usepackage{ifpdf}
\usepackage{subfigure}
\usepackage{tikz}
\usepackage[compat=1.1.0]{tikz-feynman}
\usepackage{slashed}
\usetikzlibrary{arrows,shapes}
\usetikzlibrary{trees}
\usetikzlibrary{matrix,arrows} 				
\usetikzlibrary{positioning}				
\usetikzlibrary{calc,through}				
\usetikzlibrary{decorations.pathreplacing}  
\usepackage{pgffor}							

\usetikzlibrary{decorations.pathmorphing}	
\usetikzlibrary{decorations.markings}
\tikzset{
	vector/.style={decorate, decoration={snake}, draw},
	fermion/.style={draw=black, postaction={decorate}}, 
	scalar/.style={dashed,draw=black, postaction={decorate}}}
\tikzstyle{block} = [draw, rectangle, 
minimum height=3em, minimum width=6em]
\usepackage{xcolor}
\definecolor{pantoneCB}{rgb}{0.6, 0., 0.}
\hypersetup{bookmarks=true,unicode=true,pdftoolbar=true,pdfmenubar=true,
	pdffitwindow=false,
	pdfnewwindow=true,colorlinks=true,allcolors=pantoneCB}
\usepackage{booktabs}
\usepackage{amssymb}
\usepackage{amsfonts}
\usepackage{epsf}
\usepackage{rotating}
\usepackage{graphicx}
\usepackage{amsmath}
\usepackage{fancyhdr}
\usepackage{lineno}
\usepackage{babel}
\usepackage{graphics}
\usepackage{pstricks}
\usepackage{color}
\usepackage{multirow}
\usepackage{float}
\usepackage{lineno}
\usepackage{mathtools}
\usepackage{stackengine}
\newcommand{\be}{\begin{equation}}
\newcommand{\ee}{\end{equation}}
\newcommand{\bea}{\begin{eqnarray}}
\newcommand{\eea}{\end{eqnarray}}
\def\beq{\beq\begin{aligned}}
\def\eeq{\end{aligned}\eeq}

\def\p{\textbf{p}}

\catcode`@=12

\preprint{CQUeST-2026-0804}

\title{SuperWIMPs in the Kination Era : Imprints on $N_{\rm eff}$ and CMB Spectral Distortions}
\author[a]{Genevi\`{e}ve B\'{e}langer,}
\author[b,c]{Abhishek Roy}

\affiliation[a]{LAPTh, CNRS, USMB, 9 Chemin de Bellevue, 74940 Annecy, France}
\affiliation[b]{Center for Quantum Spacetime, Sogang University, 35 Baekbeom-ro, Mapo-gu, Seoul, 121-742, South Korea}
\affiliation[c]{Department of Physics, Sogang University, 35 Baekbeom-ro, Mapo-gu, Seoul, 121-742, South Korea}
\emailAdd{belanger@lapth.cnrs.fr}
\emailAdd{abhishek@sogang.ac.kr}

\abstract
{
We investigate how the late-time decays of unstable particles into dark matter and neutrinos can leave measurable imprints on both cosmological observables and collider experiments. We study this mechanism within the scotogenic model, where the lightest singlet fermion $N_1$ serves as the dark matter candidate and is produced via the late decay of the lightest inert doublet scalar via the superWIMP mechanism.  By incorporating a kination-dominated era in the early Universe, the altered expansion rate allows to match the observed $N_1$ relic density in parameter regions that would be underabundant in standard cosmology. The energy exchange between the injected neutrinos and photons resulting from the late-time decay can produce CMB spectral $\mu$-distortions, detectable shifts in the effective number of relativistic species $N_{\rm eff}$, or both simultaneously. Concentrating on the lightest component of inert doublet scalars with masses between $200$ and $500~\text{GeV}$, we find that these cosmological signals are accompanied by complementary signatures accessible at the LHC. Although our analysis is derived within the scotogenic framework, our findings apply more broadly to scenarios in which the unstable particle undergoes late-time out-of-equilibrium decay to produce dark matter and non-thermal neutrinos, offering how future CMB measurements and collider searches can jointly probe dark matter production and the early Universe's thermal history.

}
\begin{document} 

\maketitle
\flushbottom

\section{Introduction}

Dark matter (DM) remains one of the major open questions in modern physics, and understanding its nature is a central goal of  particle physics, astrophysics and cosmology. 
Several mechanisms  have been proposed for dark matter production in the early Universe, ranging from thermal freeze-out~\cite{Scherrer:1985zt,Srednicki:1988ce} and freeze-in~\cite{McDonald:1993ex,Hall:2009bx} to non-thermal production mechanisms involving decays of heavier particles or primordial fields~\cite{Feng:2010gw}. These different scenarios can leave characteristic imprints on the evolution of the Universe and, in particular, on the primordial plasma. Big Bang Nucleosynthesis (BBN) and the Cosmic Microwave Background (CMB) therefore provide powerful cosmological probes of dark matter through its impact on the early-Universe thermal history. 
During BBN, taking place within the first few minutes after the Big Bang, dark matter particles can affect the expansion rate of the Universe  modify the thermal evolution of the primordial plasma or   inject energetic particles through annihilations and decays. These effects can alter the primordial abundances of light elements. In particular, the effective number of neutrino species, $N_{eff}$ is strongly constrained from BBN~\cite{Fields:2014uja,AtacamaCosmologyTelescope:2025nti}.
 The cosmic microwave background provides many probes of dark matter and new particles albeit at a different time, the recombination epoch. Not only does  precise measurements of the CMB allow to determine the dark matter relic density, but
 DM annihilation deposits electromagnetic energy in the
primordial plasma and impacts the anisotropies of the cosmic microwave background, 
the measurements of the CMB thus constrain s-wave dark matter below approximately 10 GeV~\cite{Slatyer:2015jla}. 
 Moreover energy injection after neutrino decoupling in the early Universe contributes to $N_{eff}$ a quantity that is also strongly constrained by the CMB. 
 In addition  energy injection  in the primordial plasma from DM annihilations
can  induce deviations from the pure black-body spectrum of the CMB. When the energy injection occurs at large redshifts, $\mu-$distortion  is the dominant effect
while at lower redshifts y-distortion dominates ~\cite{Chluba:2025wxp,Kogut:2019vqh}. Both types of distortions result in important constraints from FIRAS~\cite{Mather:1993ij,Fixsen:1996nj,Bianchini:2022dqh}.
Although $N_{eff}$ is a powerful observable, deviations from the predictions of the standard model can arise from different sources, hence the importance of combining different cosmological observables to break degeneracies. 
 It was shown in ~\cite{Li:2025clq} that a particle decaying into neutrino pairs could lead both to a deviation in the number of effective neutrinos as well as creating CMB spectral distortions, and that 
 the two signatures were  complementary and their correlation could provide information on the new physics creating these effects.

In this work, we revisit the impact of neutrinos induced from the dark sector on $N_{eff}$ and $\mu$-distortion of the CMB. 
We consider scenarios where dark matter interactions are feeble since those are particularly challenging for particle and astrophysics probes of dark matter. 
In particular, we investigate a class of scenarios where the  hidden sector contains a new particle that decays with a long lifetime into neutrino and dark matter. 
As a concrete example, we work within the scotogenic model which provides a natural motivation for a dark matter that is feebly interacting~\cite{Ma:2006km,Molinaro:2014lfa}. 
The scalar sector of the model is the one of the inert doublet model, and the lightest scalar  (hence the next-to-lightest odd particle, NLOP)
decays into a fermionic DM and a neutrino. Within this framework, we revisit the calculation of both $N_{eff}$ and $\mu$-distortion working under the assumption that the 
 lifetime of  the next-to-lightest odd particle, $A^0$, is in the range potentially relevant for inducing deviations in those two cosmological observables. This  inevitably leads to the conclusion that DM is essentially created by the superWIMP mechanism, that is the
DM is produced through the late decays of $A^0$ a  particle that was initially in thermal equilibrium with the primordial plasma~\cite{Feng:2003uy,Feng:2003xh}.
Since $Y_{A^0}$, the yield of $A^0$, is inversely proportional to the pseudoscalar mass, we concentrate on scenarios where the scalar is not too heavy (a few hundreds GeV at most) in order to maximise the effect on $N_{eff}$ and $\mu$-distortion which both depend on $Y_{A^0}$. Moreover, this mass range offers better prospects for testing at collider, providing a further complementarity between cosmological and particle physics probes.
As is well known in the inert doublet model where $A^0$ is the DM,  masses  in the range 100-500 GeV, lead to too low a relic  abundance. Since in the superWIMP mechanism the DM abundance is directly linked to that of the NLOP,  this model can provide a 
viable DM candidate only if it is realised in non-standard cosmological scenarios with low reheating temperature, notably in kination~\cite{Spokoiny:1993kt,Joyce:1996cp,Belanger:2025ack}. Our study will therefore highlight the connection between cosmological probes,  dark matter and non-standard cosmology. 

The paper is organized as follows. Section~\ref{sec:model} contains the description of the scotogenic model; Section~\ref{sec:DMobs} describes the calculation of the dark matter density; in Section~\ref{Sec:Como_Sig} we discuss  the cosmological signatures $N_{eff}$ and CMB $\mu$-distortion. The dependence of these two observables on the free parameters of the model is highlighted as well. Section~\ref{Sec:collider} contains a brief discussion of the collider signatures associated with the model, while Section~\ref{sec:conclusion} contains our conclusions. More details on the computation of cosmological observables  can be found in the Appendix~\ref{appendix}. 
\begin{table}[]
	\centering
	\begin{tabular}{|c|c|c|c|c|c|c|}
		\hline \phantom{XXXXXXXX} & \phantom{X} $L_i$ \phantom{X} & \phantom{X}$\ell_{R_i}$\phantom{X} & \phantom{X} $H$ \phantom{X} & \phantom{X} $\Phi$ \phantom{X} & \phantom{X} $N_i$ \phantom{X}  \\
		\hline \hline
		{\bf  $SU(2)_L$} & {\bf  $2$} & {\bf  $1$} & {\bf  $2$} & {\bf  $2$} & {\bf  $1$}  \\
		\hline
		{\bf  $U(1)_Y$} & {\bf  $-1/2$} & {\bf  $-1$} & {\bf  $1/2$} & {\bf  $1/2$} & {\bf  $0$} \\
		\hline
		$\mathbb{Z}_2$ & $+$ & $+$ & $+$  & $-$ & $-$ \\
		\hline
	\end{tabular}
	\caption{Particle content and their charge assignments under various symmetry groups.}
	\label{tab:particle-content}
\end{table}

\section{The Model}
\label{sec:model}
The Scotogenic model, which is also referred to as the radiative seesaw model, is a simple extension of the Standard Model (SM) by a $SU(2)_{L}$ Higgs doublet $\Phi$ with hypercharge $Y = 1/2$ along with three SM gauge singlet fermions, $N_i\,(i=1,2,3)$. 
Furthermore, the model features a discrete $\mathbb{Z}_2$ symmetry under which all new fields are odd while the SM ones are even. The $\mathbb{Z}_2$ symmetry forbids tree-level flavour-changing neutral current as well as the tree-level Dirac mass for neutrino thus generating neutrino mass radiatively. It also ensures the stability of the lightest odd particle, yielding a viable DM candidate which can be either the lightest odd scalar or fermion In this work, we will consider the lightest fermion, $N_1$, as a dark matter candidate. The colourless particle content and their associated quantum number under the SM gauge group  and $\mathbb{Z}_2$ symmetry 
is summarised in Table~\ref{tab:particle-content}. The renormalizable gauge-invariant terms in the Lagrangian that generate neutrino masses are given by,
\begin{equation}
-\mathcal{L}_{\text{Y}} \supset Y^{\ell}_{\alpha i} \overline{L}_{\alpha} H \ell_{R_i} + Y^{\nu}_{i\alpha } \,  \overline{N}_i \tilde{\Phi}^\dagger L_{\alpha} + \frac{1}{2} M_{N_k} \overline{N_k^c}N_k +\text{h.c.},
\label{eq:LagYL}
\end{equation}
where $\tilde{\Phi}=i\sigma_2\Phi^*$, $L=(\nu_{\ell L},\ell_L)^T$, and the singlet fermions mass matrix, $M_{N_k}$, is taken to be diagonal. The invariant scalar potential can be written as,
\begin{align}
V =  &\mu_1^2 H^\dagger H + \mu_2^2 \Phi^\dagger \Phi + \frac{\lambda_1}{2} (H^\dagger H)^2 + \frac{\lambda_2}{2} (\Phi^\dagger \Phi)^2 + \lambda_3 (H^\dagger H)(\Phi^\dagger \Phi) + \lambda_4 (\Phi^\dagger H) (H^\dagger \Phi )+ \nonumber \\  &\frac{\lambda_5}{2} \left[ (\Phi^\dagger H)^2 + \text{h.c} \right].\label{eq:potential}
\end{align}
For definiteness, all parameters in the above are assumed to be real, and $\mu_2^2 > 0$ is imposed to prevent spontaneous breaking of the $Z_2$ symmetry. Working in the unitary gauge, the scalar field about the vacuum can be written as,
\begin{align}
H=
\begin{pmatrix}
0\\
(v+h)/\sqrt{2}
\end{pmatrix} \,\,\text{and}\,\,
\Phi=
\begin{pmatrix}
H^+\\
(H^0+iA^0)/\sqrt{2}
\end{pmatrix},
\label{eqmass}
\end{align}
where $v= 246\, \text{GeV}$ is the SM Higgs vacuum expectation value (VEV), and $h$ is SM Higgs boson with mass $M_h\simeq 125\,\rm{GeV}$. The masses of new scalar particles can be expressed as, 
\begin{align}
M_{H^{+}}^{2}&=\mu_{2}^{2}+\frac{1}{2}\lambda_{3}v^{2},\\
M_{H^0}^{2}&=\mu_{2}^{2}+\frac{1}{2}\lambda_{L}v^{2},\\
M_{A^0}^{2}&=\mu_{2}^{2}+\frac{1}{2}\bar{\lambda}_{L}v^{2},
\end{align}
where we have defined
\begin{align}
\lambda_{L}=\left(\lambda_{3}+\lambda_{4}+\lambda_{5}\right)\,\,\,\text{and} \,\,\, \bar{\lambda}_{L}= \left(\lambda_{3}+\lambda_{4}-\lambda_{5}\right).
\end{align}
The scalar potential in Eq.~\ref{eq:potential} contains seven independent parameters. As we have assumed $H$ to be the SM Higgs doublet, $\mu_1$ and $\lambda_1$ are fixed by the observed Higgs mass and electroweak data. The remaining five parameters can therefore be traded for the following set of five independent parameters,
\begin{align}
\{\lambda_2,\, \bar{\lambda}_L,\,M_{H^{\pm}},\,M_{H^0},\,M_{A^0} \}.
\end{align}
It is important to note that $\lambda_2$ does not affect the scalar mass spectrum at tree level, but rather plays a pivotal role in ensuring the stability of the electroweak vacuum. Therefore, throughout our work, we fix $\lambda_2 = 0.1$ for the sake of simplicity. In addition, we assume $M_{A^0} < M_{H^0}$; since $\lambda_5 = (M_{H^0}^2 - M_{A^0}^2)/v^2$, this implies $\lambda_5 > 0$ \footnote{This choice does not affect the dark matter phenomenology}.
\par The $\mathbb{Z}_2$ symmetry forbids the tree-level seesaw contribution to neutrino masses. Instead, neutrino masses are generated at one-loop via the exchange of dark fermions and scalars, with the neutrino mass matrix  given by \cite{Mandal:2021yph,Ma:2006km,DeRomeri:2022cem,Vicente:2014wga,Chun:2023vbh}:
\begin{align}
\label{eq:numass}
(m_\nu)_{ij} &
= \sum_{k=1}^3 \frac{Y^\nu_{ik} \, Y^\nu_{kj} M_{N_k}}{32 \pi^2}\left[ \frac{M_{H^0}^2}{M_{H^0}^2-M_{N_k}^2} \log \frac{M_{H^0}^2}{M_{N_k}^2} -  \frac{M_{A^0}^2}{M_{A^0}^2 - M_{N_k}^2} \log \frac{M_{A^0}^2}{M_{N_k}^2} \right],\\
& \equiv (Y_\nu^T \Lambda Y_{\nu})_{ij},
\end{align}
where the $\Lambda$ matrix is defined as $\Lambda=\text{diag}(\Lambda_1,\Lambda_2,\Lambda_3)$, with
\begin{align}
\label{eq:Lambda}
\Lambda_k
=  \frac{M_{N_k}}{32 \pi^2}\left[ \frac{M_{H^0}^2}{M_{H^0}^2-M_{N_k}^2} \log \frac{M_{H^0}^2}{M_{N_k}^2} -  \frac{M_{A^0}^2}{M_{A^0}^2 - M_{N_k}^2} \log \frac{M_{A^0}^2}{M_{N_k}^2} \right].
\end{align}

To  explore 
the implications of neutrino data, we adopt the Casas-Ibarra parametrization~\cite{Casas:2001sr} for the Yukawa coupling matrix:
\begin{align}
Y^\nu = \sqrt{\Lambda}^{-1} R \sqrt{\hat{m}_\nu} U^{\dagger}_{\rm lep},
\label{eq:Ynu}
\end{align}
Here, $R$ is a general $3 \times 3$ complex orthogonal matrix, $U_{\rm lep}$ denotes the Pontecorvo-Maki-Nakagawa-Sakata (PMNS) matrix~\cite{ParticleDataGroup:2014cgo,Granelli:2024nmo} responsible that diagonalizes the neutrino mass matrix, and $m_\nu$ is the diagonal matrix of light neutrino masses. In our analysis, we set $R = \mathbb{I}$ for simplicity. In this limit, Eqn.~\ref{eq:Ynu} simplifies, and can be written as,
\begin{equation}\label{Eqn:Similified_Ynu}
    Y^{\nu}_{i\alpha}=\sqrt{\frac{m_{\nu,i}}{\Lambda_i}}\left(U^{\dagger}_{\rm lep}\right)_{i\alpha}.
\end{equation}
It is worth noting that, in the limit $m_{\nu,1}\to 0$, where $m_{\nu,1}$ denotes the lightest active neutrino mass, the corresponding $N_1$ decouples from the theory. We further assume normal ordering for the neutrino masses and fix the oscillation parameters to their current best-fit values from Ref.~\cite{deSalas:2020pgw}. 
It is important to bring to the reader's attention that the model is subject to various experimental bounds and theoretical conditions for consistency, including
\begin{itemize}
    \item Vaccum Stability~\cite{Belyaev:2016lok}
    \item Perturbativity~\cite{Belanger:2022qxt}
    \item Perturbative Unitarity~\cite{Arhrib:2012ia}
    \item Global Minimum Condition and Charge-Breaking Vacua~\cite{Ginzburg:2010wa,Belyaev:2016lok}
    \item LEP data~\cite{Lundstrom:2008ai,Pierce:2007ut}
    \item Charge Lepton Flavour Violation~\cite{Toma:2013zsa,Lindner:2016bgg,MEG:2016leq}
    \item Electroweak precision data~\cite{Barbieri:2006dq}
\end{itemize}
To keep the discussion focused on the main analysis of this work, we refer the reader to the following references for a detailed discussion of these constraints \cite {Belyaev:2020wok,Roy:2025moo,Avila:2025qsc,Chun:2026tsq}. In our analysis, we impose all the constraints listed above.

\section{Dark Matter Relic Density}
\label{sec:DMobs}
In this work, we consider that the lightest $Z_2$-odd fermion $N_1$ is a feebly interacting  DM candidate. Our primary objective is to investigate the impact of a low reheating temperature on dark matter phenomenology in this setup.  We assume a mass hierarchy of the form $M_{N_1}<M_{A^0}<M_{H^0},M_{H^\pm}\ll M_{N_{2,3}}$, under which $N_1$ serves as a viable DM candidate, while $A^0$ is the next-to-lightest odd particle (NLOP). It is important to emphasize that, due to the chosen mass hierarchy, $N_{2,3}$ undergo early chemical decoupling from the SM thermal bath and have a negligible impact on $N_1$ production. Therefore, in our work, we fix $M_{N_{2,3}} = 10\,\mathrm{TeV}$, while keeping the remaining masses below the TeV scale. Furthermore, the relic density of $N_1$ has two main contributions: the freeze-in production and the late decay of the NLOP, commonly known as the super-WIMP mechanism. Quantitatively, it can be written as~\cite{Belanger:2022gqc},
\begin{align}\label{Eqn:relic_N1}
    \Omega_{N_1}h^2&=\Omega^{\rm{freeze-in}}_{N_1}h^{2}+\Omega^{\rm{super-wimp}}_{N_1}h^{2}\nonumber\\
    &=\Omega^{\rm{freeze-in}}_{N_1}h^{2}+\frac{M_{N_1}}{M_{A^0}}\Omega_{A^0}h^2.
\end{align}
We will discuss the importance of each contribution in a low reheating temperature scenario in the subsequent sections.

To capture the effects of a low reheating temperature, we investigate the evolution of $N$ and $A^0$ by solving coupled Boltzmann equations in a cosmological background, dominated initially by a decaying inflaton field. We express the number densities in terms of the comoving yield $z = n a^3$, where $n$ denotes the number density and $a$ is the cosmological scale factor, for the dark matter candidate $N$ and the NLOP $A^0$. The corresponding Boltzmann equations are given by:
\begin{align}
    \frac{dz_{N_1}}{da}&=\sum_{X=A^0,H^0,H^{\pm}}\frac{\Gamma_{X\to N_1 \nu\left(l^{\pm}\right)}}{Ha}\bar{z}_{X},\nonumber \\
    \frac{dz_{A^0}}{da} &= -\frac{\langle \sigma_v \rangle_{eff}}{Ha^4} \left(z_{A^0}^2 - \bar{z}_{A^0}^2\right)-\frac{\Gamma_{A^0\to N_1 \nu}}{Ha}z_{A^0},
    \label{Eqn:Beq1_N1}
\end{align}
 where $\bar{z} = n_{\text{eq}} a^3$ denotes the comoving equilibrium yield. The equilibrium number density $n_{\text{eq}}$ is given by~\cite{Kolb:1988aj}
\begin{equation}
    n_{\text{eq}}(T) = \frac{g}{2\pi^2} M_1^2 T K_2\!\left(\frac{M_1}{T}\right),
\end{equation}
where $g$ accounts for the internal degrees of freedom of the dark sector particles, and $K_2$ is the modified Bessel function of the second kind. Furthermore, $H$ denotes the Hubble expansion rate, $\langle \sigma_v \rangle_{eff}$ is the effective thermally averaged annihilation cross-section $A^0$, which includes all pair annihilation as well as co-annihilation processes involving $H^0$ and $H^{\pm}$. $\Gamma_{X\to Nl^{\pm}(\nu)}$ is the partial decay width of $X(= H^0,\, A^0,\, H^{\pm})$ to $N_1$. Note that the production of $N_1$ through $2\to 2$ processes, either via annihilation or scattering of thermal bath particles, is generally subdominant to its production through bath particle decays. This is because the corresponding interaction rates are suppressed by higher powers of $Y_{\alpha 1}^\nu$, additional phase-space factors, and propagator suppression. The DM relic density is obtained from  
\begin{equation}
    \Omega_{N_1}h^2=2.742\times10^8 M_{N_1}Y^0_{N_1},
\end{equation}
where $Y^0_{N_1}(=\frac{n_0}{s_{0}})$ is the asymptotic value of the DM yield
and $ n _ {0} $ and $s_0\simeq2.69\times10^{-9}\,\rm{cm^{-3}}$ are the present-day number density and entropy density, respectively. The DM relic density has been precisely determined from observations of the cosmic microwave background (CMB) by the Planck satellite~\cite{Planck:2018vyg}:
\begin{equation}
    \Omega^{\rm obs}_{N_1} h^2 = 0.120 \pm 0.0012.
\end{equation}
Throughout this work, we require the predicted relic density to lie within the $2\sigma$ range of this measurement. The solution of Eqn.~\ref{Eqn:Beq1_N1} depends sensitively on the cosmological background, which is determined by the energy densities of the inflaton ($\rho_\phi$) and radiation ($\rho_R$). The following equations govern their evolution,~\cite{Bernal:2019mhf, Garcia:2020wiy,Maldonado:2019qmp,Arcadi:2024jzv}:
\begin{align}
&\frac{d\rho_\phi}{dt}+3(1+w)H\rho_{\phi}=-\Gamma_{\phi}\rho_{\phi},\\
&\frac{d\rho_R}{dt}+4H\rho_{R}=\Gamma_{\phi}\rho_{\phi},
\label{eqn: backround_evolution1}
\end{align}
where $\Gamma_\phi$ corresponds to the inflaton decay width, and $w$ is the equation-of-state (EoS) parameter. Furthermore, the Hubble expansion rate is sourced by both $\rho_R$ and $\rho_{\phi}$,
\begin{equation}
H^2 =H^2_{SM}+H^2_{\phi}=\frac{8\pi\rho_\phi}{3M_P^2} + \frac{8\pi\rho_R}{3M_P^2}.
\end{equation}
Here $M_P = 1.2 \times 10^{19}\,\text{GeV}$ is the Planck mass, while the SM energy density is given by,
\begin{align}
    \rho_{SM}(T)=\rho_R(T)=\frac{\pi^2}{30}g_{eff}(T)T^4,
\end{align}
where $g_{eff}$ corresponds to number of degree of freedom that contribute to $\rho_R$.

\par In the kination scenario that we consider here, corresponding to $w=1$, the energy density of $\phi$ redshifts faster than that of SM radiation. As a result, the Universe can become radiation dominated without requiring $\phi$ to decay into SM particles. In other words, the evolution is insensitive to $\Gamma_{\phi}$; hence, we set it to be zero. Thus, we can express the energy density of $\phi$ as~\cite{Barman:2021ugy,Bernal:2024yhu}
\begin{align}
    \rho_{\phi}(T)=\rho_R(T_{RH})\left(\frac{T}{T_{RH}}\right)^6,
    \label{Eqn:rho_phi_Vs_T}
\end{align}
where $T_{RH}$ denotes the reheating temperature of the Universe, which is set by the condition $\rho_\phi(T_{RH})=\rho_R(T_{RH})$. It can be expressed in terms of the Hubble expansion rates at the onset of reheating as,
\begin{align}
    T_{RH}\simeq T_{MAX}\left[\frac{H_{SM}(a_I)}{H_{\phi}(a_I)}\right].
\end{align}
Here, $a_I=1$ corresponds to the scale factor at the onset of reheating and $T_{MAX}$ corresponds to the maximum temperature of the thermal bath and is given by,
\begin{align}
    T_{MAX}\simeq\left[\sqrt{\frac{90}{\pi^2g_{eff}}}M_P H_{SM}(a_I) \right]^{1/2}.
\end{align}
Note that the observations of the cosmic microwave background constrain $H_\phi(a_I) \lesssim 4 \times 10^{-6} M_P$~\cite{BICEP:2021xfz} and to satisfy BBN constraint we impose $T_{RH}>T_{BBN}\simeq4\,\rm MeV$~\cite{Kawasaki:2000en,Hannestad:2004px,deSalas:2015glj,DEramo:2017gpl}. Throughout this work, we fix $T_{MAX}=10^6\,\rm{GeV}$. To summarise, the list of independent parameters relevant to our analysis are,
\begin{align}
\boxed{
\bar{\lambda}_L,\,M_{H^{\pm}},\,M_{H^0},\,M_{A^0},\,M_{N_1},\,\tau_{A^0},T_{RH}
}
\end{align}
where $\tau_{A^0} = 1/\Gamma_{A^0}$ corresponds to the lifetime of $A^0$. The total decay width $\Gamma_{A^0}$ can be written as,
\begin{align}
\Gamma_{A^0}
&=\sum_{\alpha}\frac{|Y_{1\alpha}|^2M_{A^0}}{32\pi}
\left(1-\frac{M_{N_1}^2}{M_{A^0}^2}\right)^2,\nonumber\\
&=\frac{m_{\nu,1}}{\Lambda_1}\frac{\sum_{\alpha}(U^{\dagger}_{\rm lep})^2_{1\alpha}M_{A^0}}{32\pi}
\left(1-\frac{M_{N_1}^2}{M_{A^0}^2}\right)^2,
\end{align}
 where the Yukawa couplings $Y^\nu_{1\alpha}$ are related to the light active neutrino masses $m_{\nu,1}$ through the Casas-Ibarra parameterization, as shown in Eqn.~\ref{Eqn:Similified_Ynu}. In this work, we impose that $A^0$ decays after neutrino decoupling from the thermal bath, thus requiring that  $\tau_{A^0}$ lies in the range $10^3$--$10^{10}\,\mathrm{secs}$,  corresponding to  $Y^\nu_{1\alpha}$ of $\mathcal{O}(10^{-16}\text{--}10^{-12})$. We will discuss the phenomenological implications  in subsequent sections.  The two direct consequences of the assumed range for $\tau_{A^0}$ are that (i) $N_1$ is a non-thermal DM and (ii) in the  kination scenario that we consider, $N_1$  is produced by the superWIMP mechanism since the production of DM via the FIMP mechanism  generally requires $Y^\nu_{i\alpha} \gtrsim 10^{-9}$~\cite{Silva-Malpartida:2023yks,Choudhury:2026aao,Amiri:2025ras,Bernal:2024ndy}, a value much larger than the one corresponding  to the  chosen lifetime range. Note that even assuming the instantaneous inflaton decay approximation with $T_{RH}$ marking the onset of radiation domination, $N_1$ yield from the FIMP mechanism is given by~\cite{Longas:2026yxq},
\begin{align}
Y_{N_1} \sim \sum_{X=H^0,A^0,H^\pm}\frac{0.0871 g_{X}}{g_s \sqrt{g_*}}\frac{M_P \Gamma(X \to N_1) }{M^2_{X}} \left(\frac{M_{X}}{T_{RH}} \right)^{5/2} e^{-M_{X}/T_{RH}},
\label{Eqn:Ins_FIMP_N1}
\end{align}
from which  we can infer  the exponential suppression of the FIMP yield when $M_{N_1} \gg T_{RH}$ as well as a suppression  due to the very narrow width of X
Therefore, this ensures that the super-WIMP mechanism gives the primary contribution to  $N_1$ production in our assumed $\tau_{A^0}$ range, i.e., $10^3$--$10^{10}\,\mathrm{secs}$. Consequently, Eqn.~\ref{Eqn:relic_N1} simplifies to,
 \begin{align}\label{Eqn:relic_N1_final}
    \Omega_{N_1}h^2\approx \frac{M_{N_1}}{M_{A^0}}\Omega_{A^0}h^2.
\end{align}
\begin{figure}[]
\centering
\mbox{\subfigure[]{\includegraphics[width=0.5\textwidth]{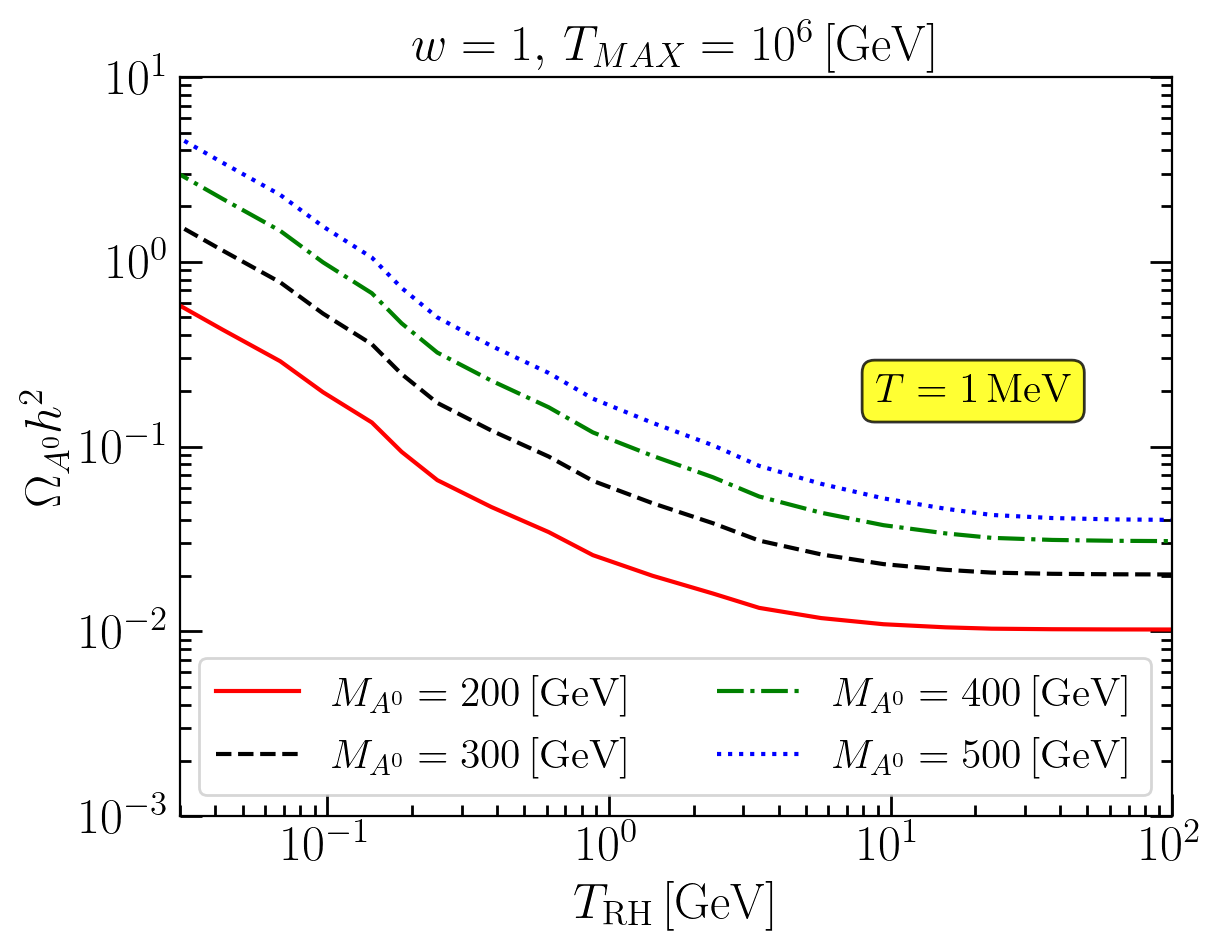}\label{Fig:BP_1a}}
\subfigure[]{\includegraphics[width=0.5\textwidth]{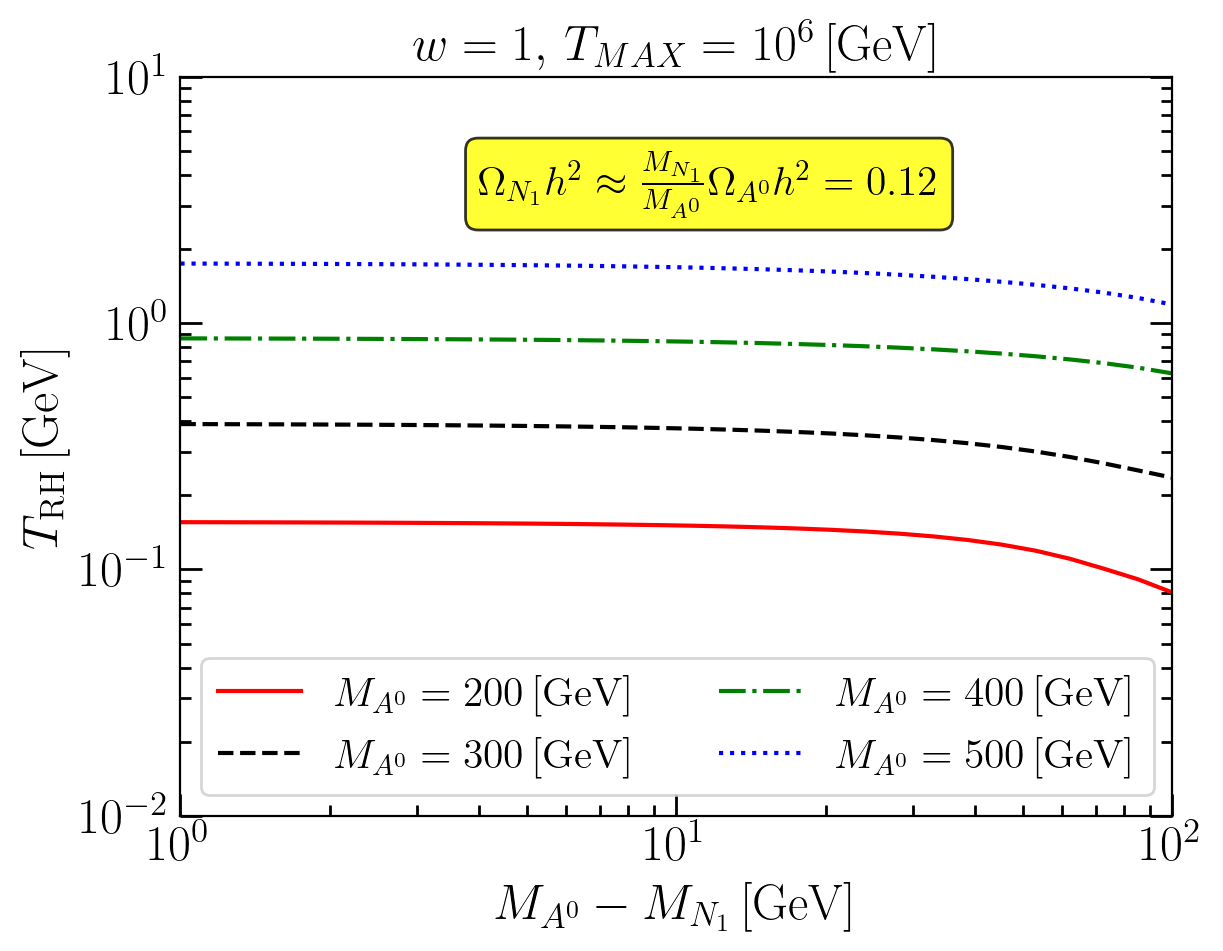}\label{Fig:BP_1b}}
}
\caption{(a) $\Omega_{A^0}h^2$  vs $T_{RH}$, the red, black, green and blue lines correspond to $M_{A^0}=200,\,300,\,400,$ and $500\,\rm GeV$ respectively.
(b) Correlation between  $T_{RH}$ and $M_{A^0}-M_{N_1}$. We have fixed $M_{H^0}-M_{A^0}=2\,\rm{GeV},\, M_{H^\pm}-M_{H^0}=10\,\rm{GeV},\,\text{and}\,\bar{\lambda}_{L}=0$. All curves satisfy the neutrino oscillation best-fit data.} 
\label{Fig:Benchmark_Point1}
\end{figure}
Thus, reproducing $\Omega_{N_1}h^2\approx0.12$, requires $\Omega_{A^0}h^2\gtrsim0.12$, given that $M_{N_1}\lesssim M_{A^0}$. Since $\Phi$ is charged under the SM electroweak gauge group, all components of $\Phi$ thermalize with the SM thermal bath, and their abundances follow the standard freeze-out mechanism. The lightest component of $\Phi$, i.e., $A^0$, annihilates efficiently in the mass range $\sim 120\,\rm GeV$ to $\sim 500\,\rm GeV$, rendering $\Omega_{A^0}h^2\lesssim 0.12$ ~\cite{Belyaev:2018ext} in the standard cosmological scenario. In particular, even when the Higgs-portal coupling $\bar{\lambda}_L$ which governs the $A^0$ annihilation via the SM Higgs boson is set to zero, $A^0$ primarily annihilates to SM gauge bosons driven by the fixed SM gauge coupling and remains very efficient. Consequently, a sufficiently large $A^0$ mass is required to suppress its (co)-annihilation rate, and we find that $\Omega_{A^0}h^2\simeq 0.12$ is achievable only for $M_{A^0}\gtrsim 500\,\rm GeV$~\cite{LopezHonorez:2006gr,Goudelis:2013uca,Belyaev:2016lok,Belanger:2025ack} 
In Fig.~\ref{Fig:BP_1a}, we show that this conclusion can change in a non-standard cosmological scenario with kination~\cite{Belanger:2025ack}.  For low $T_{RH}$, the additional $\rho_{\phi}$ component, which scales as $T^6$ (see Eqn.~\ref{Eqn:rho_phi_Vs_T}), drives the rapid expansion of the Universe and enables early chemical decoupling of $A^0$ from the thermal bath during the kination era.  Thus, $A^0$ annihilation becomes less efficient, increasing its relic abundance. At high $T_{RH}$, $A^0$ decouples during the radiation-dominated phase and results are similar to those obtained in  the standard cosmological scenario. For a fixed value of $T_{RH}$, we expect that $\Omega_{A^0}h^2$ increases when $M_{A^0}$ increases. This behavior follows from the decrease in $\langle\sigma_v\rangle_{eff}$, as $M_{A^0}$ increases. Because the dominant annihilation channel remains unchanged, the reduced annihilation rate leads to a larger $\Omega_{A^0}h^2$. 
As $M_{A^0}> M_{N_1}$, $A^0$ decays out of equilibrium in the late Universe and produces the $ N_1$ number density. In Fig.~\ref{Fig:BP_1b}, we show the parameter space in the $(M_{A^0}-M_{N_1})$ -$T_{RH}$ plane, considering $M_{A^0}$ in the range $200-500\,\rm GeV$ and the mass splitting  $(M_{A^0}-M_{N_1})$ varying from $1\,\rm GeV$ to $100\,\rm GeV$.  We find that $T_{RH}$ decreases as the mass splitting increases. This behavior can be understood from the fact that a larger mass splitting corresponds to a smaller $N_1$ mass. Thus, to reproduce the observed DM relic abundance requires a larger $\Omega_{A^0}h^2$, which in turn corresponds to a smaller $T_{RH}$.

It is also important to bring to the readers' attention that the injection of energetic neutrinos can initiate secondary electroweak and hadronic cascades and alter primordial light element abundances --- specifically deuterium ($\text{D}$), helium-3 ($^3\text{He}$), helium-4 ($^4\text{He}$), and lithium ($^6\text{Li}, ^7\text{Li}$) --- through spallation and photodisintegration~\cite{Akita:2024ork,Ema:2014ufa,Kanzaki:2007pd,Hambye:2021moy,Acharya:2020gfh,Bianco:2025boy,GrimbaumYamamoto:2025nai}. Hence, BBN constraints from light-element observations, along with observables $\Delta N_{\text{eff}}$ and CMB spectral distortions, constrain both the lifetime $\tau_{A^0}$ and the yield of long-lived particles, $Y_{A^0}$. We  adopt a conservative bound following Ref.~\cite{Bianco:2025boy,GrimbaumYamamoto:2025nai} and impose $Y_{A^0} \lesssim 10^{-11}$ for lifetimes $\tau_{A^0} \in [10^3, 10^{10}] \text{secs}$, assuming $E_\nu^\star < 90\text{ GeV}$~\footnote{Note that the bound is derived from the decay of NLOP $\phi$, i.e., $ \phi \to \nu\bar{\nu}$ decays, rather than $A^0 \to N_1\bar{\nu}$; a rigorous derivation of the limits for the latter channel is beyond the scope of this work and is left for future investigation.}. Furthermore, requiring $N_1$ to account for the observed DM relic abundance, we find that the resulting BBN constraint limits the allowed mass splitting between $A^0$ and $N_1$. In particular, $M_{A^0}=200\,\text{GeV}$ and $M_{N_1}\lesssim 100\,\text{GeV}$ leads to $Y_{A^0}\gtrsim 10^{-11}$.

\section{Cosmological Signatures: $N_{\rm eff}$ and CMB $\mu$-distortions}\label{Sec:Como_Sig}
The early Universe provides a useful means of investigating new physics by examining its effects on the evolution of the cosmic plasma and the radiation content. The injection of electromagnetic energy or additional light particles, such as Majorons or non-thermal neutrinos produced through the decay or annihilation of BSM particles during the late epoch of the Universe, can alter the thermal history of neutrino sectors or produce the departure of the background photon spectrum from a perfect blackbody distribution, producing observable features in $\Delta N_{\rm eff}$, CMB spectral distortions, or both. While $\Delta N_{\rm eff}$ reflects the changes in radiation energy density as compared to the standard cosmological prediction, spectral distortions occur when energy injection causes the photon distribution to deviate from an ideal blackbody. Thus the two observables examine different parts of the thermal history and can give complementary information on non-standard processes in the early Universe~\cite{Li:2025clq}. To compute these observables, we start with the evolution of the neutrino distribution function after neutrino decoupling, including both cosmological redshifting and additional neutrino injection from the $A^0 \to N_1 \bar{\nu}$ process (where $N_1$ is a viable DM candidate). The Boltzmann equation for the evolution of the injected neutrino distribution function can be written as,
\begin{align}
\frac{d f_{\nu}^{\rm inj}}{d  t} =\frac{\partial f_{\nu}^{\rm inj}}{\partial t}-H|\p|\frac{\partial f_{\nu}^{\rm inj}}{\partial |\p|}=\mathcal{C}_\nu[f^{\rm inj}_\nu] \,.
\end{align}
To determine the distribution function $f_{\nu}^{\rm inj}$ of the injected neutrinos, we solve the Boltzmann equation under the following approximations: 

\begin{enumerate}
 \item
\begin{itemize}
\item[] We assume homogeneous and isotropic distribution functions for both the neutrinos and DM. In the absence of a preferred spatial or momentum direction, the distribution functions depend only on the magnitude of the momentum, $p$, and the cosmic time, $t$. We can therefore write $f=f(t,|p|)$, reducing the phase-space dependence of the Boltzmann equation.
\end{itemize}

\item 
\begin{itemize}
\item[] We neglect the scattering of the injected neutrinos with background electrons for temperatures below neutrino decoupling. After neutrino decoupling, at $T\lesssim \mathcal{O}(1\,\mathrm{MeV})$, the neutrinos are no longer in thermal equilibrium with the electromagnetic plasma. Moreover, after $e^+e^-$ annihilation, the Universe charge neutrality fixes the net electron abundance to the proton abundance, $n_{e^-}-n_{e^+}\simeq n_p\simeq\eta_B n_\gamma$, with $\eta_B\sim \mathcal{O}(10^{-10})$. The resulting scarcity of background electrons suppresses the neutrino--electron scattering rate, allowing the corresponding collision term to be neglected to a good approximation.
\end{itemize}
\item
\begin{itemize}
\item[] We neglect inverse-decay processes and Pauli-blocking effects.  The distribution function of the injected particles satisfies $f\ll1$. Thus, the Pauli-blocking factor can be taken as $(1-f)\simeq1$, and the inverse-decay contribution can be neglected.
\end{itemize}
\end{enumerate}

We first calculate the contribution to the neutrino phase space distribution from the decay process $A^0\rightarrow N_1+\nu$, characterized by the decay width $\Gamma_{A^0}$, which contributes as a source term to the neutrino phase space distribution. The collision term can be written as,
\begin{equation}
    \mathcal{C}_{\nu}[f^{\rm inj}_\nu]
=\frac{1}{2E_{\nu}}\int \left(\prod_{i=A^0,N_1} \frac{d^3p_i}{(2\pi)^32E_i} \right) (2\pi)^{4} \delta^{4}(p_{A^0}-p_{N_1}-p_{\nu}) |\mathcal{M}|^{2} f_{A^0} 
\, .
\end{equation}
Since $A^0$ is non-relativistic during its decay, we take
\begin{align}
    E_{A^0}&\simeq M_{A^0},\nonumber\\
     f_{A^0}&=(2\pi)^3\delta^3(p_{A^0})n_{A^0},\nonumber
\end{align}
the collision term becomes~\footnote{We refer the reader to Appendix~\ref{Appendix:Collision_Term_Calc} for the detailed derivation of the collision terms.},
\begin{align}
    \mathcal{C}_{\nu}[f^{\rm inj}_\nu]=\frac{2\pi^2\Gamma_{A^0}Y_{A^0}c_sT^3}{E^{\star}_{\nu}E_{\nu}}e^{\eta(1-x^2)}\delta(E^\star_\nu-E_\nu),
\end{align}
where $c_s\equiv\frac{2\pi^2}{45}g_s\simeq1.47$, $x=T_{\nu\rm dec}/{T}$, with the neutrino decoupling temperature taken to be $T_{\nu\rm dec}\simeq 1\,\rm{MeV}$~\cite{Escudero:2018mvt}, and $E^\star_\nu$ corresponds to the injected neutrino energy in the rest frame of $A^0$,
\begin{equation}
    E^\star_\nu=\frac{M^2_{A^0}-M^2_{N_1}}{2M_{A^0}}.
    \label{Eqn:Enu_COM}
\end{equation}
Furthermore, $\eta$ measures the ratio of $A^0$ decay rate with respect to Hubble rate at $T_{\nu\rm dec}$ and can be written as,
\begin{equation}
    \eta=\frac{\Gamma_{A^0}M^\star_p}{2T^2_{\nu\rm dec}},
\end{equation}
where $M_p^\star=\frac{0.6M_p}{\sqrt{g_\rho}}
\simeq 4\times10^{18}\,\mathrm{GeV}$ assuming that $g_\rho\simeq g_s\simeq3.36$ remains approximately constant after neutrino decoupling, which is a good approximation for $T\lesssim1\,\rm{MeV}$~\cite{Husdal:2016haj}. The yield $Y_{A^0}$ encodes the abundance of the parent particle $A^0$ at the onset of its decay and, in general, depends on the underlying model parameters as well as the assumptions about the early Universe. In particular, $Y_{A^0}$ depends on the production mechanism of $A^0$ and can depend on quantities such as its mass, couplings, reheating temperature $T_{RH}$, and the inflaton equation of state $w$. 
We define  the dimensionless variables, 
\begin{align}
 r=\frac{|{\bf E}_\nu|}{T}\,,  \quad r_{\star}= \frac{E^{\star}_\nu}{T_{\nu\rm dec}},
\end{align}
and impose the initial condition $f_{\nu}^{\rm inj}(t_{\nu,\rm dec},|\p|)=0$. The distribution function of the injected neutrinos is obtained by integrating the collision term and is given by~\footnote{The detailed derivation of the injected neutrino distribution function is provided in Appendix~\ref{Appendix:Dis_Func_Calc}.}
\begin{align}
f_{\nu}^{\rm inj}(t, |\p|)  
=\frac{4\pi^2 \eta Y_{A^0}c_s}{rr^2_{\star}}e^{\eta\left(1-\frac{r^2}{r^2_\star}\right)}\Theta(r,r_\star,x).
\label{Eqn:Distri_func}
\end{align}
The Heaviside step-function $\Theta(r,r_\star,x)\equiv \theta(r-r_\star)\theta(x r_\star-r)$ specifies the allowed momentum range of the injected neutrinos. The first factor reflects the monochromatic momentum fixed by the decay of the nonrelativistic $A^0$, while the second accounts for the redshift of this momentum between the time of production and the time of observation.
\subsection{Energy density and contribution to $N_{eff}$}
The effective number of relativistic species, $N_{\rm eff}$, offers another important means of investigating the thermal history of the early Universe; in the standard cosmological model, neutrinos decouple from the electromagnetic plasma when the temperature is of the order of a few MeV, and thereafter the annihilation of electrons and positrons transfers entropy predominantly to the photon bath, resulting in~\cite{Lesgourgues:2012uu}, 
\begin{equation}
    \frac{T_\nu}{T_\gamma}
    =
    \left(\frac{4}{11}\right)^{1/3}.
\end{equation}
The neutrino energy density can then be expressed in terms of the photon energy density as
\begin{equation}
    \rho_\nu
    =
    \frac{7}{8}
    \left(\frac{4}{11}\right)^{4/3}
    N_{\rm eff}\,\rho_\gamma.
\end{equation}
In the Standard Model, non-instantaneous neutrino decoupling and finite-temperature QED effects predict $N_{\rm eff}\simeq 3.0443$~\cite{Escudero:2025kej}.

The prediction can be altered if there are any further relativistic degrees of freedom or non-standard energy injections into the neutrino or electromagnetic sectors. The relevant sequence can be illustrated as follows
\begin{equation}
\resizebox{0.99\textwidth}{!}{$
\boxed{
\text{New energy/particle injection}
\;\longrightarrow\;
\nu,\;e^\pm,\;\gamma
\;\longrightarrow\;
\text{energy transfer and decoupling}
\;\longrightarrow\;
\Delta\rho_{\rm rad}
\;\longrightarrow\;
\Delta N_{\rm eff}.
}
$}\nonumber
\end{equation}
Hence, the various processes that took place near or after neutrino decoupling can cause a change in the relative energy densities of the neutrino and photon backgrounds, resulting in a modification of the total radiation density which can be expressed in terms of \(\Delta N_{\rm eff}\). The non-thermal neutrinos injected from $A^0$ decay after neutrino decoupling provide an additional contribution to the total radiation energy density. This contribution can be quantified in terms of the injected neutrino energy density, which is obtained by integrating the injected phase-space distribution over momentum:
\begin{align}
\delta\rho_{\nu}^{\rm inj}=2\int \frac{d^{3} p_\nu}{(2\pi)^{3}} E_{\nu} f_{\nu}^{\rm inj}(E_\nu,T)
\label{Eqn:Neff}
\end{align}
where the factor of $2$ in Eqn.~\ref{Eqn:Neff} accounts for the equal contribution from neutrinos and antineutrinos. The corresponding contribution to the effective number of relativistic species is then obtained by comparing the injected neutrino energy density with the photon energy density,
\begin{align}
\Delta N_{\rm eff}&=\frac{8}{7}\left(\frac{11}{4}\right)^{4/3}\left(\frac{\delta\rho_{\nu}^{\rm inj}}{\rho_{\gamma}}\right)\nonumber\\
&=6.672\int_{r_\star}^{xr_\star} dr r^3f^{\rm inj}_{\nu}(r,x)
\label{Eqn:Neff_final}
\end{align}
\begin{figure}[]
\centering
\mbox{
\subfigure[]{\includegraphics[width=0.5\textwidth]{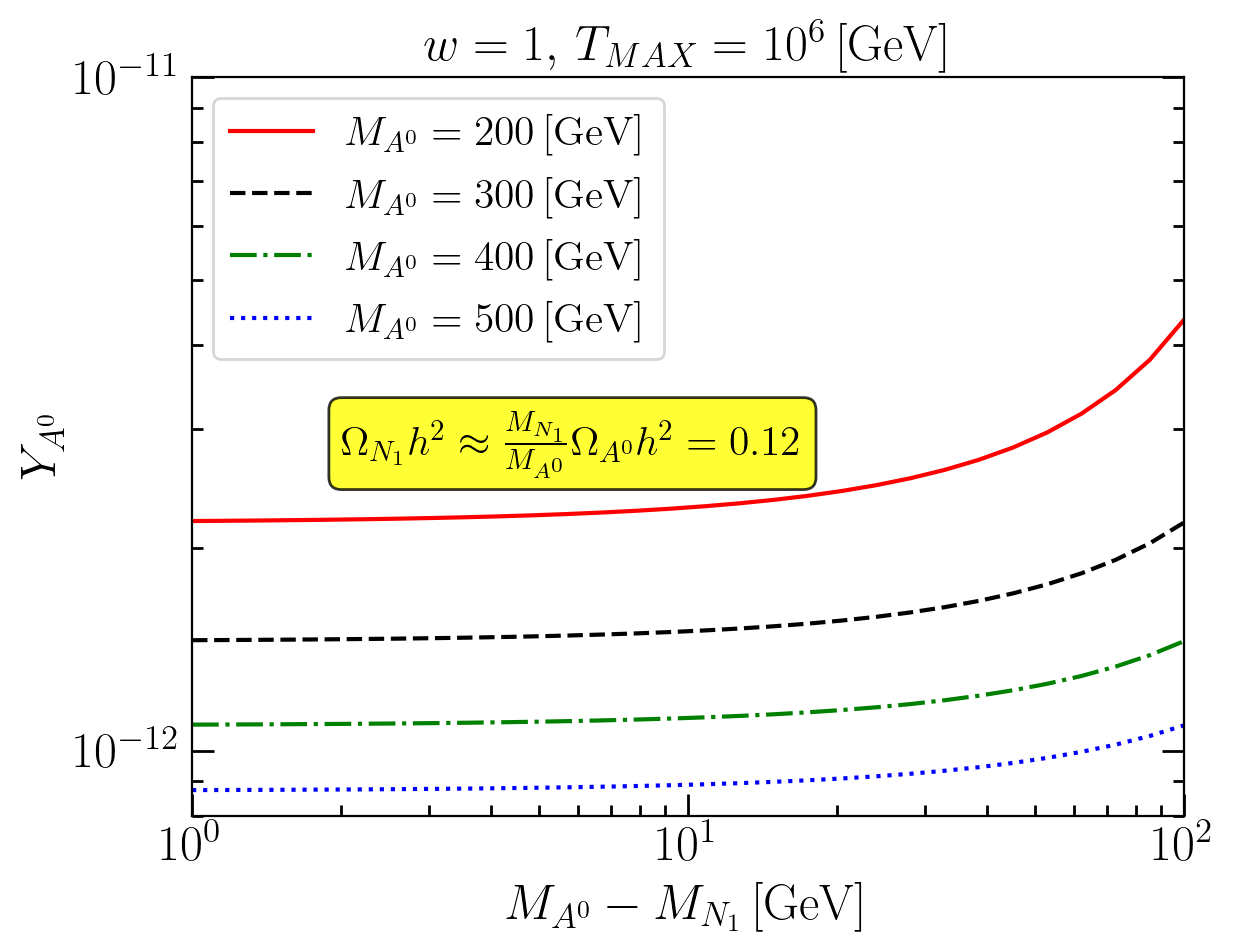}\label{Fig:Neff_BP2a}}
\subfigure[]{\includegraphics[width=0.5\textwidth]{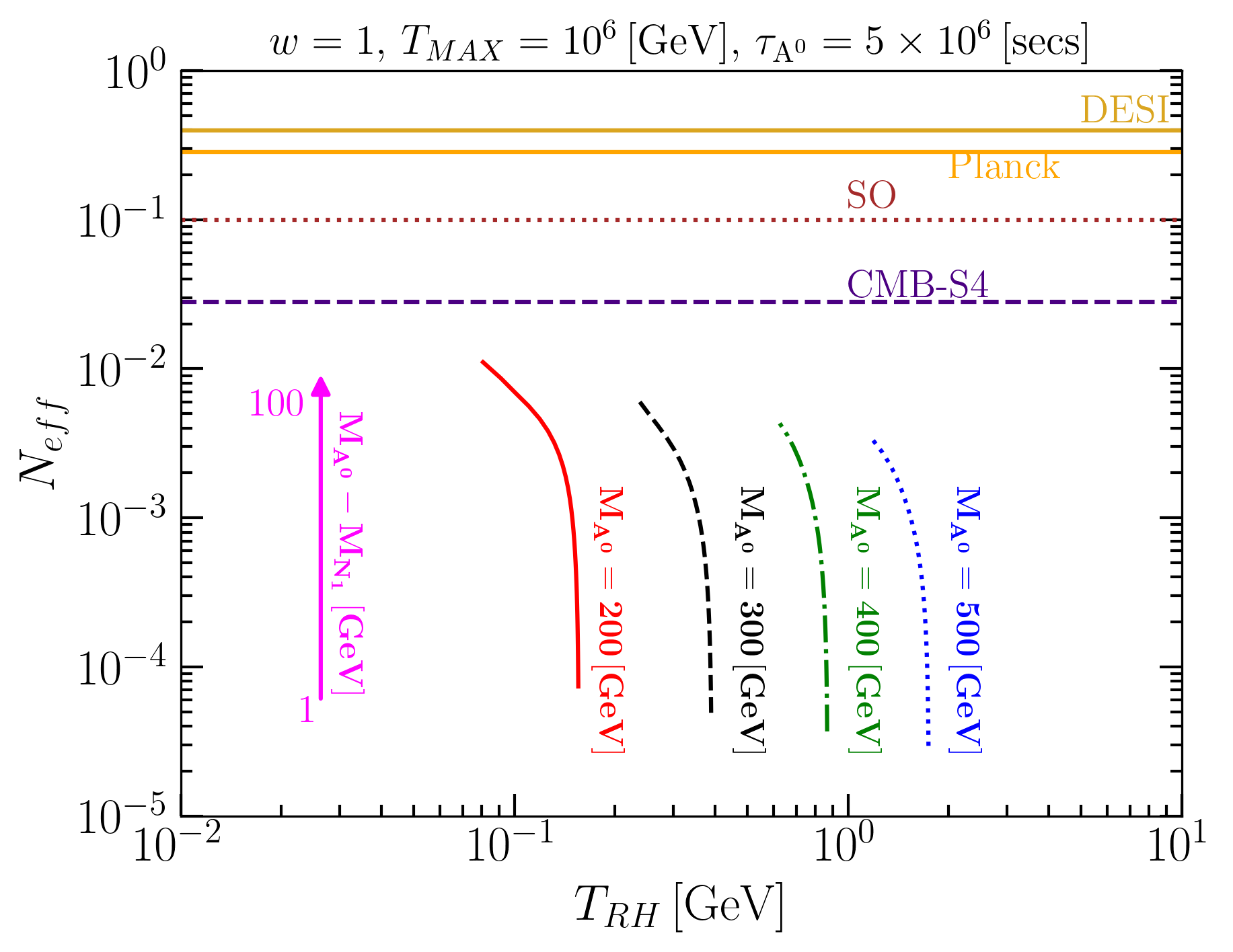}\label{Fig:Neff_BP2b}}
}
\mbox{
\subfigure[]{\includegraphics[width=0.5\textwidth]{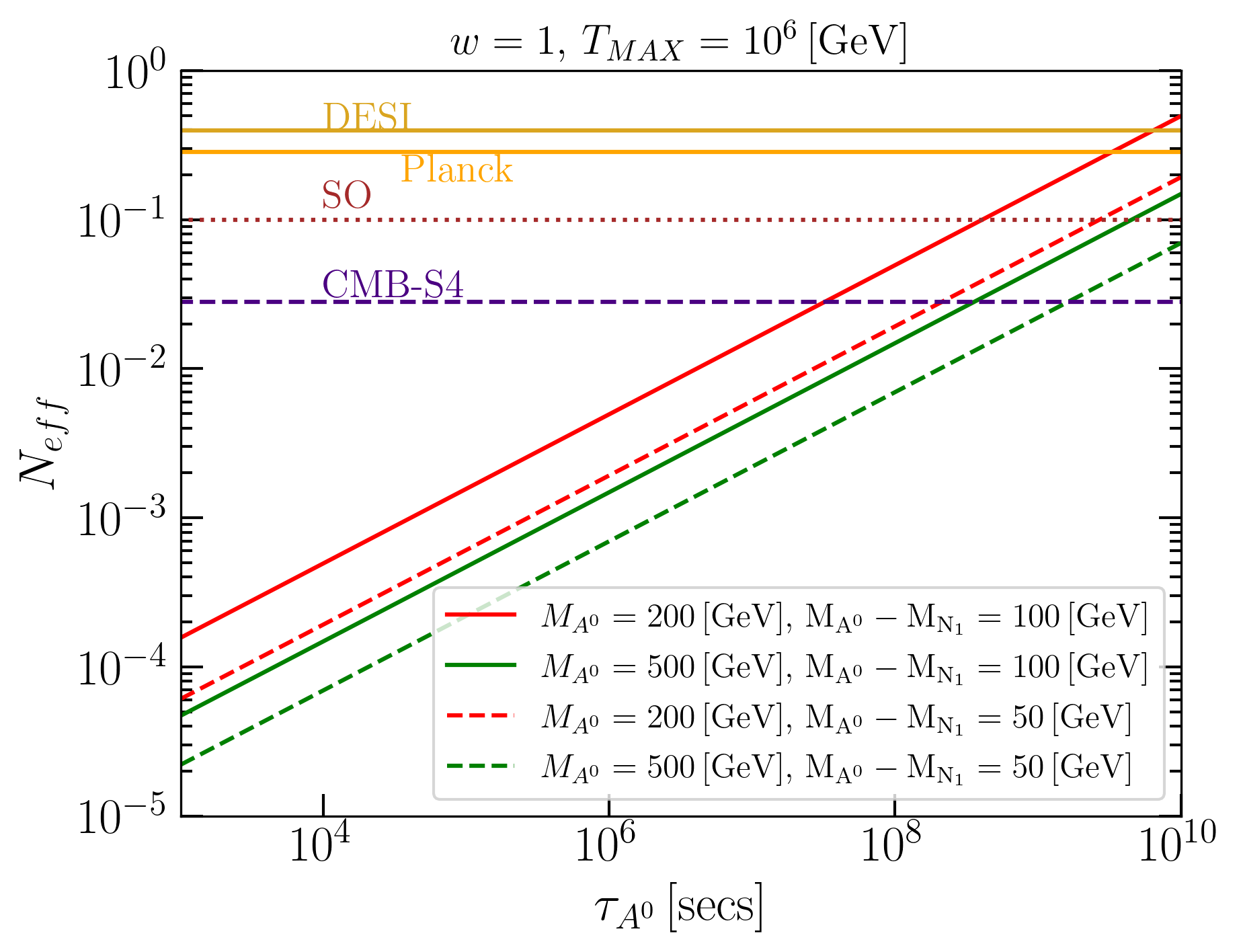}\label{Fig:Neff_BP2c}}
}
\caption{
(a) Correlation between $Y_{A^0}$ at the neutrino decoupling temperature, $T_{\nu,\rm dec}\approx 1\,\rm MeV$, and the mass splitting $M_{A^0}-M_{N_1}$ for  different values of $M_{A^0}$.  
(b) $N_{\rm eff}$ as function of $T_{RH}$ for different  values of $M_{A^0}$ and $M_{A^0}-M_{N_1}$,  assuming $\tau_{A^0}=5\times10^6\,\rm{secs}$.
(c) Variation of $N_{\rm eff}$ with $\tau_{A^0}$ for mass splittings $M_{A^0}-M_{N_1}=50\,\rm GeV$ and $100\,\rm GeV$ and  $M_{A^0}=200\,\rm GeV$ and $500\,\rm GeV$. All the points satisfy the relic density limit within 2$\sigma$ and the best-fit neutrino oscillation data.}
\label{Fig:Neff_BP2}
\end{figure}
Once $A^0$ has decayed to $N_1$ and $\nu$ after thermal decoupling from the thermal bath, $\Delta N_{\rm eff}$ approaches a constant value, and we use its asymptotic value at recombination ($T\sim 0.1\,\mathrm{eV}$) in our analysis. In Fig.~\ref{Fig:Neff_BP2a}, we show how the yield $Y_{A^0}$ at neutrino decoupling varies with the mass splitting $M_{A^0}-M_{N_1}$ for different values of $M_{A^0}$. 

As $M_{A^0}$ decreases from $500\,\rm GeV$ to $200\,\rm GeV$, the yield increases as $Y_{A^0}\propto 1/M_{A^0}$. For a fixed $M_{A^0}$, $Y_{A^0}$ also increases with the mass splitting, since a lighter $N_1$ requires a larger $A^0$ yield to reproduce the observed DM relic abundance. Since $f_{\nu}^{\rm inj}\propto Y_{A^0}$ (see Eqn.~\ref{Eqn:Distri_func}), this implies that $\Delta N_{\rm eff}\propto Y_{A^0}$. Thus, from Fig.~\ref{Fig:Neff_BP2b}, which shows the variation of $N_{\rm eff}$ in the $N_{\rm eff}-T_{RH}$ plane, we can see that $N_{\rm eff}$ is always larger for $M_{A^0}=200\,\rm GeV$ than for larger values of $M_{A^0}$. For a fixed $M_{A^0}$, moving from bottom to top along the corresponding curve, $M_{A^0}-M_{N_1}$ increases from $1$ to $100\,\rm GeV$, as indicated by the magenta arrow. In this direction, the energy of the injected neutrinos, $ E_\nu^\star$ (see Eqn.~\ref{Eqn:Enu_COM}), also increases, increasing $N_{\rm eff}$.
However, we see that all points fall below the $N_{\rm eff}$ forecast limit for the assumed lifetime of $\tau_{A^0}=5\times10^6\,\rm secs$. In Fig.~\ref{Fig:Neff_BP2c}, we show the correlation between $\tau_{A^0}$ and $N_{\rm eff}$ for $M_{A^0}-M_{N_1}=100\,\rm GeV$ and $50\,\rm GeV$, corresponding to $M_{A^0}=200\,\rm GeV$ and $500\,\rm GeV$, respectively.
We have fixed $M_{H^0}-M_{A^0}=2\,\rm{GeV},\, M_{H^\pm}-M_{H^0}=10\,\rm{GeV},\,\text{and}\,\bar{\lambda}_{L}=0$. We observe that $N_{\rm eff}$ increases with $\tau_{A^0}$. This happens because the decay occurs later, when the background radiation density has decreased due to cosmic expansion. Meanwhile, the energy density of the non-relativistic parent particle $A^0$ redshifts more slowly than radiation ($\rho_{A^0}\propto a^{-3}$, whereas $\rho_{\rm rad}\propto a^{-4}$). Consequently, its energy density becomes larger relative to the background radiation at the time of decay, leading to a larger contribution to $N_{\rm eff}$. Note that the current $2\sigma$ observational bounds on $\Delta N_{\rm eff}$  are  $\Delta N_{\rm eff}<0.395$~\cite{DESI:2024mwx} from DESI and  $\Delta N_{\rm eff}<0.285$ from PLANCK~\cite{Planck:2018vyg}. The upcoming Simons Observatory (SO) experiment will reach $\Delta N_{\rm eff}= 0.1$ at $2\sigma$~\cite {SimonsObservatory:2018koc}, and CMB-S4 will reach $\Delta N_{\rm eff}=0.028$~\cite{CMB-S4:2026mge}.  We can see that, for large $\tau_{A^0}$, all configurations are within the reach of future $N_{\rm eff}$ measurements while $M_{A^0}=200\,\rm GeV$ is already excluded when 
$\tau_{A^0}\gtrsim 4\times 10^9$.

\subsection{CMB $\mu$ Distortion}
The emergence of a \(\mu\)-type spectral distortion can be understood from the interplay between energy injection, Compton scattering, and photon-number-changing processes in the early Universe. Consider an epoch in which $A^0$ decays into $N_1$ and an injected neutrino,
\begin{equation}
A^0 \rightarrow \bar{N_1}+\nu_{\rm inj}.\nonumber
\end{equation}
The injected neutrinos can undergo pair annihilation~\footnote{We neglect final-state radiation in the process $\nu_{\rm inj}+\bar{\nu}_{\rm inj}\rightarrow e^-+e^++\gamma$. The interaction rate is suppressed because it is higher order in the electromagnetic coupling than the corresponding $\nu_{\rm inj}+\bar{\nu}_{\rm inj}\rightarrow e^-e^+$ process and can therefore be neglected to a very good approximation.},
\begin{equation}
\nu_{\rm inj}+\bar{\nu}_{\rm inj}\rightarrow e^++e^-.\nonumber
\end{equation}
The energy of the injected neutrinos is set by the decay kinematics and can be much greater than the thermal plasma temperature. The annihilation process is kinematically possible only when the injected neutrinos have sufficient energy to produce an $e^+e^-$ pair;  this condition can be written as
\begin{equation}
E_{\nu,\rm inj}\gtrsim m_e.
\end{equation}

In addition to annihilation among the injected neutrinos, they can also co-annihilate with the background neutrinos,
\begin{equation}
\nu_{\rm inj}+\bar{\nu}_{\rm bg}
;(\bar{\nu}_{\rm inj}+\nu_{\rm bg})
\rightarrow e^++e^-.\nonumber
\end{equation}
In this case, the background neutrino has a typical energy set by the neutrino temperature,
\begin{equation}
E_{\nu,\rm bg}\sim T.
\end{equation}
Because the background neutrino energy is much less than the injected neutrino energy, a greater energy has to be injected in order to reach the threshold for $e^+e^-$ production; this condition can be expressed as,
\begin{equation}
E_{\nu,\rm inj}E_{\nu,\rm bg}\gtrsim m_e^2,\nonumber
\end{equation}
which gives
\begin{equation}
E_{\nu,\rm inj}\gtrsim\frac{m_e^2}{T}.
\label{Eqn:Coann_threshold}
\end{equation}
The produced electrons and positrons can transfer their energy to the background photon bath through inverse Compton scattering~\footnote{For relativistic $e^\pm$ pairs, inverse Compton scattering with the background photons provides the dominant energy-loss mechanism and proceeds much faster than the Hubble expansion before recombination~\cite{Blumenthal:1970gc,Padmanabhan:2005es,Chen:2003gz}.},
\begin{equation}
e^\pm+\gamma_{\rm bg}\rightarrow e^\pm+\gamma.\nonumber
\end{equation}
The resulting energetic photons subsequently scatter with the background electrons and positrons through Compton scattering,
\begin{equation}
e^{\pm}_{\rm bg}+\gamma\rightarrow e^\pm+\gamma,\nonumber
\end{equation}
redistributing the injected energy within the electromagnetic plasma. Since Compton scattering conserves the total number of photons, it cannot by itself restore a Planck distribution when the photon energy density has changed.

At sufficiently high redshifts, photon-number-changing reactions are also efficient. The most important processes are double Compton scattering~\cite{Chluba:2013kua,Chluba:2025wxp},
\begin{equation}
    e^-+\gamma \leftrightarrow e^-+\gamma+\gamma,\nonumber
\end{equation}
and bremsstrahlung,
\begin{equation}
    e^-+p \leftrightarrow e^-+p+\gamma.\nonumber
\end{equation}
These reactions can create or absorb photons and therefore allow the photon number to adjust to the increased energy density. Consequently, any energy injection occurring at very early times is efficiently thermalized and the photon distribution returns to a blackbody spectrum.

As the Universe expands and cools, the photon-number-changing processes become inefficient before Compton scattering does. An energy injection during this intermediate epoch is then redistributed by Compton scattering without a corresponding adjustment of the photon number density. The photon distribution consequently approaches a Bose--Einstein form,
\begin{equation}
    f_\gamma(E)
    =
    \frac{1}{e^{E/T+\mu}-1},\nonumber
\end{equation}
with a nonzero chemical potential \(\mu\). This gives rise to the \(\mu\)-type CMB spectral distortion.
The overall sequence can therefore be summarized schematically as,
\begin{equation}
\resizebox{0.99\textwidth}{!}{$
\boxed{
\text{Energy injection}
\;\longrightarrow\;
e^\pm/\gamma
\;\xrightarrow{\rm Compton}
\text{energy redistribution}
\;\longrightarrow\;
\text{Bose--Einstein spectrum}
\;\longrightarrow\;
\mu\text{-distortion}.
}
$}\nonumber
\end{equation}
The survival of the distortion is determined by whether photon-number-changing interactions are sufficiently rapid to erase it. Thus, the \(\mu\)-distortion provides a probe of energy release during the epoch in which Compton scattering remains efficient while double Compton scattering and bremsstrahlung have become inefficient.
Later, when Compton scattering becomes inefficient, energy injection can no longer fully redistribute photon energy to establish a Bose--Einstein distribution. Instead, the distortion takes the form of a $y$-type spectral distortion, characterized by a frequency-dependent change in the CMB spectrum due to energy transfer between electrons and photons~\cite{Chluba:2013kua,Chluba:2025wxp,Barenboim:2025vme}. A detailed study of the resulting $y$-distortion and its associated constraints is beyond the scope of this work and is left for future studies.

The $\mu$ distortion can be calculated by integrating the energy transfer rate over the entire production history and can be written as~\cite{Sunyaev:1970er,Chluba:2016bvg,Li:2024xlr,Chluba:2013vsa},
\begin{align}
	\mu\approx 1.4 \int_{0}^{\infty} \mathcal{J}_\mu(T)\frac{{\rm d}\rho_\gamma/{\rm d}t}{\rho_\gamma HT}{\rm d}T,
    \label{Eqn:mu_dis}
\end{align}
where the $\mathcal{J}_\mu$ function describes the visibility of the energy injection during the epoch when the $\mu$ distortion is formed. It is given by~\cite{Chluba:2013vsa,Li:2024xlr,Burigana:1991eub,Barenboim:2025vme,Hu:1992dc,Chluba:2016bvg}
\begin{align}
\mathcal{J}_\mu(T)=e^{-(T/T_\mu)^{5/2}}\theta(T-T_{\mu y})\,,
\end{align}
where $T_\mu\approx 0.47$~keV (corresponding to $z\approx2\times 10^6$, or equivalently, $t\approx\mathcal{O}(10^6)\,\rm{secs}$) denotes the upper end of the $\mu$-distortion era, while $T_{\mu y}\approx 12$~eV (corresponding to $z=5\times 10^4$ , or equivalently, $t\approx\mathcal{O}(10^{10})\,\rm{secs}$) marks the transition from $\mu$ to $y$ distortion. The electron-positron pair system will heat the background photons and can be approximately  given by~\cite{Li:2025clq,Leng:2026iqm,Chen:2003gz,Li:2024xlr,Padmanabhan:2005es},
\begin{align}
	\frac{{\rm d}\rho_\gamma}{{\rm d}t}\approx \frac{{\rm d}\rho_{e\bar e}}{{\rm d}t}\,.
\end{align}
\begin{figure}[]
\centering
\mbox{
\subfigure[]{\includegraphics[width=0.5\textwidth]{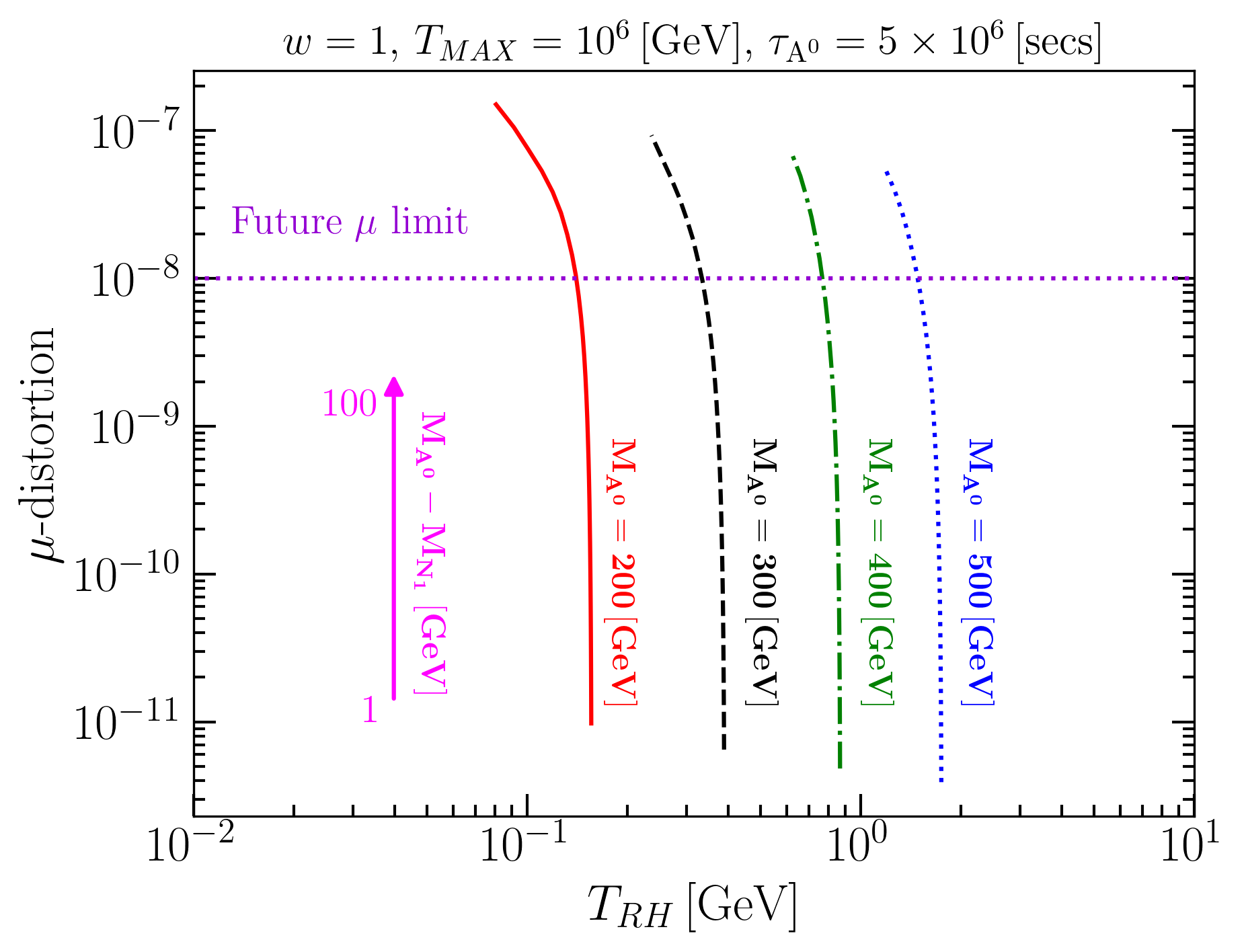}\label{Fig:mu_tau3a}}
\subfigure[]{\includegraphics[width=0.5\textwidth]{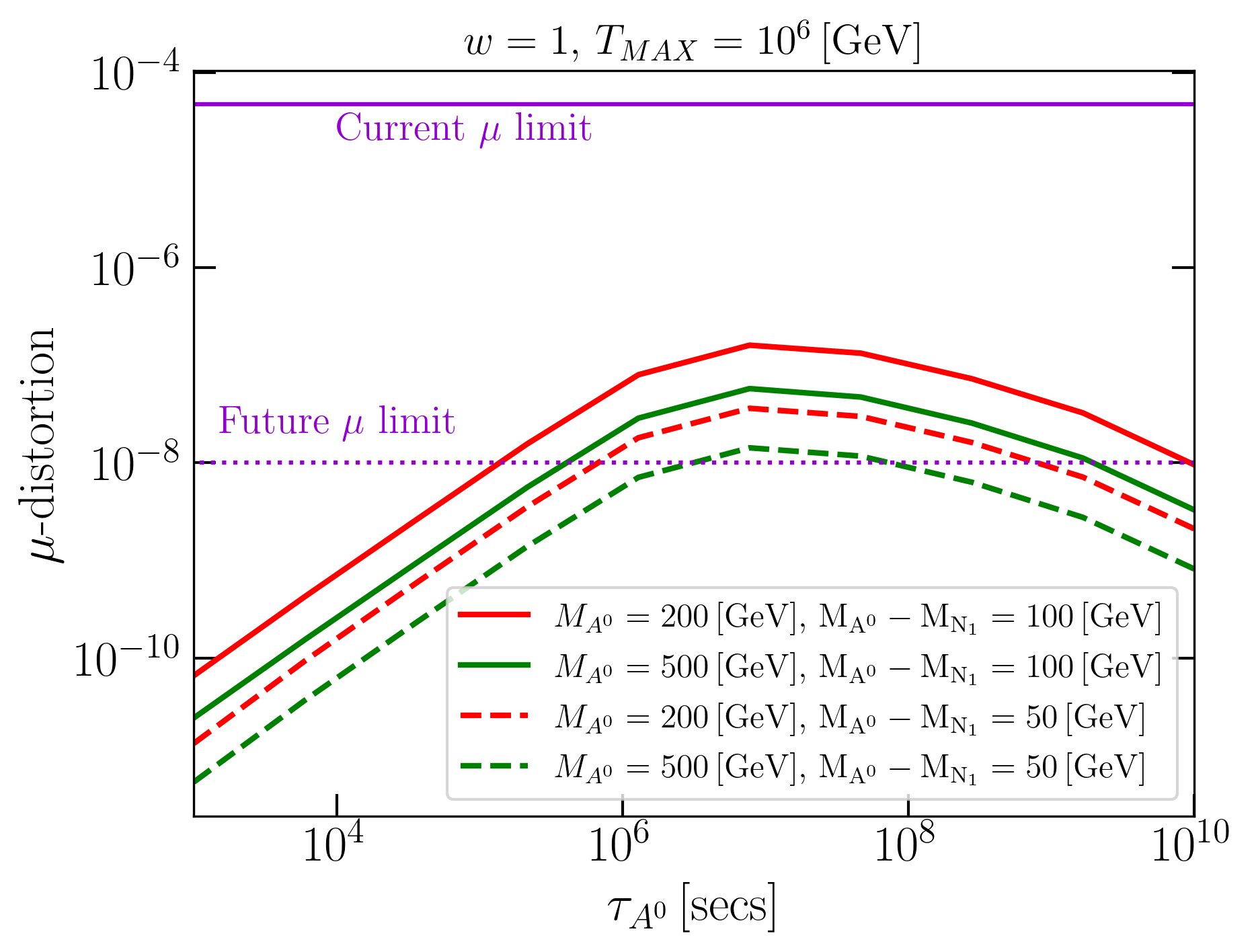}\label{Fig:mu_tau3b}}
}
\caption{(a) $\mu-$distortion dependence on $T_{RH}$ for different values of $M_{A^0}$ and of $M_{A^0}-M_{N_1}$ assuming $\tau_{A^0}=5\times10^6\,\rm{secs}$. 
(b) Variation of $\mu-$distortion  with $\tau_{A^0}$ for $M_{A^0}-M_{N_1}=50\,\rm GeV$ and $100\,\rm GeV$ and $M_{A^0}=200\,\rm GeV$ and $500\,\rm GeV$. We have fixed $M_{H^0}-M_{A^0}=2\,\rm{GeV},\, M_{H^\pm}-M_{H^0}=10\,\rm{GeV},\,\text{and}\,\bar{\lambda}_{L}=0$.  All the points satisfy the relic density limit  within $2\sigma$ and  the best-fit neutrino oscillation data.}
\label{Fig:mu_tau3}
\end{figure}
The energy transfer rate into the electron-positron pair is then given by~\footnote{We refer the reader to Appendix~\ref{Appendix:Energy_Tranfer_rate} for the detailed calculation for the energy transfer rate into an electron-positron pair.}, 
\begin{align}
	\frac{{\rm d}\rho_{e\bar e}}{{\rm d}t}
    &=\int \left(\prod_{i=\nu,\bar \nu, e,\bar e}\frac{{\rm d}^3 {\bf p}_i}{(2\pi)^3 2E_i} \right) \delta^4(p_\nu+p_{\bar \nu_{}}- p_e-p_{\bar e})|\mathcal{M}|^2_{\nu\to e} (E_e+E_{\bar e})f^{\rm inj}_\nu (f^{\rm inj}_{\bar \nu}+ f^{\rm bg}_{\bar \nu}),\nonumber\\
    &=\frac{G_{\rm F}^2}{192\pi^5}\int dE_\nu dE_{\bar \nu}(E_\nu+E_{\bar \nu})f^{\rm inj}_{\nu} (f^{\rm inj}_{\bar \nu}+f^{\rm bg}_{\bar \nu})\mathcal{K}(E_\nu,E_{\bar\nu}),
\end{align}
where $\mathcal{K}(E_\nu,E_{\bar\nu})$ can be written as,
\begin{align}
    \mathcal{K}(E_\nu,E_{\bar\nu})=\frac{64}{3}g_W\left(E^3_\nu E^3_{\bar\nu}-m^6_e \right)+8(2g_W-3)m^2_e\left(E^2_\nu E^2_{\bar\nu}-m^4_e\right),\nonumber
\end{align}
and $f^{\rm inj}_\nu$ and $f^{\rm bg }_{\nu}$ denote the non-thermal injected and thermal background neutrino distribution functions, respectively.  The processes involving one injected neutrino and one background neutrino correspond to co-annihilation and are proportional to $f^{\rm inj}_\nu f^{\rm bg }_{\nu}$. On the contrary, annihilation of two injected neutrinos gives the pair-annihilation contribution, proportional to $f^{\rm inj}_\nu f^{\rm inj}_{\nu}$ (and similarly for antineutrinos). Thus, the latter is quadratic in the injected population, whereas the co-annihilation term is linear in $f^{\rm inj}_\nu$ for a fixed thermal background. In principle, the injected neutrinos can undergo fast flavor oscillations. Depending on the temperature at which they are produced, the oscillation timescale can be much shorter than both the Hubble and the interaction timescale. The oscillation timescale is given by~\cite{Escudero:2018mvt},
\begin{align}
    t_{\rm osc}\sim \frac{T}{\Delta m^2}\sim 10^{-5}\left(\frac{T}{\rm MeV}\right).
    \label{Eqn:Time_oscillation}
\end{align}
In our work, we focus on the region of parameter space where the injected neutrino is produced after the thermal neutrinos decouple from the thermal bath, i.e., $T<T_{\nu,\rm dec}$; in this case, neutrinos undergo rapid oscillations. Therefore, we multiply Eqn.~\ref{Eqn:Time_oscillation} by a factor of 3 to compute the energy transfer rate to the electron-positron plasma. The original COBE/FIRAS limits on CMB spectral distortions ~\cite{Mather:1993ij,Fixsen:1996nj}
have been improved by reanalyses to $|\mu| < 4.7 \times 10^{-5}$~\cite{Bianchini:2022dqh} and $|y| < 8.3 \times 10^{-6}$~\cite{Sabyr:2025hwd}. New experiments are now in active development and will start taking data shortly. These include ground based arrays (COSMO~\cite{Masi:2021azs}), balloon borne missions (BISOU~\cite{Maffei:2021xur}) and a number of space based experiments (PIXIE~\cite{Kogut:2011xw, Kogut:2019vqh,Kogut:2024vbi}, SPECTER~\cite{Sabyr:2024lgg}) and FOSSIL~\cite{Aghanim:2026mii} which are all aiming to get down to $|\mu| \sim 10^{-8}$ and $|y| \sim 10^{-9}$. We take  $|\mu| \sim 10^{-8}$ as the future reach in our analysis. 
\begin{figure}[]
\centering
\mbox{
\subfigure[]{\includegraphics[width=0.5\textwidth]{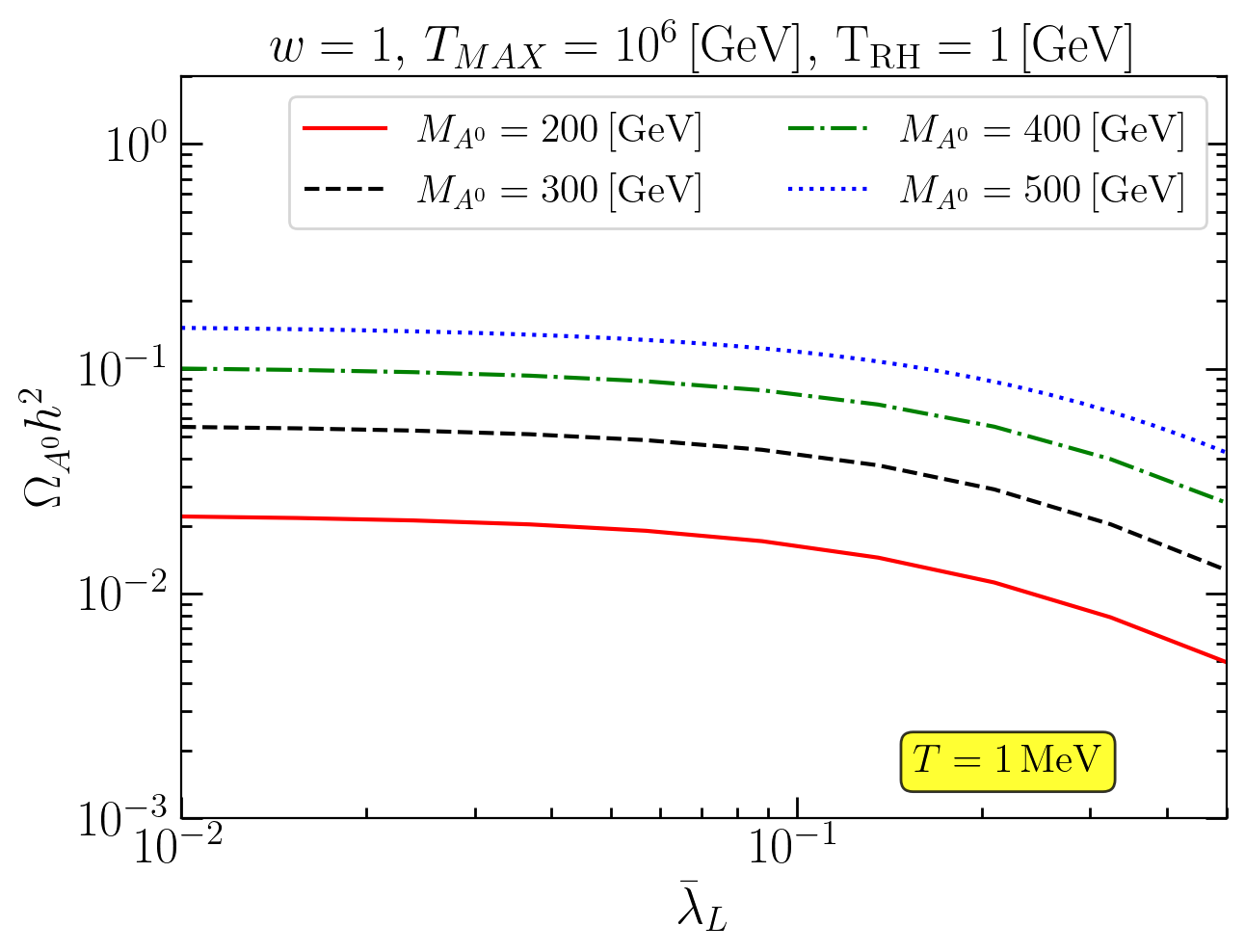}\label{Fig:laml_CMB4a}}
\subfigure[]{\includegraphics[width=0.5\textwidth]{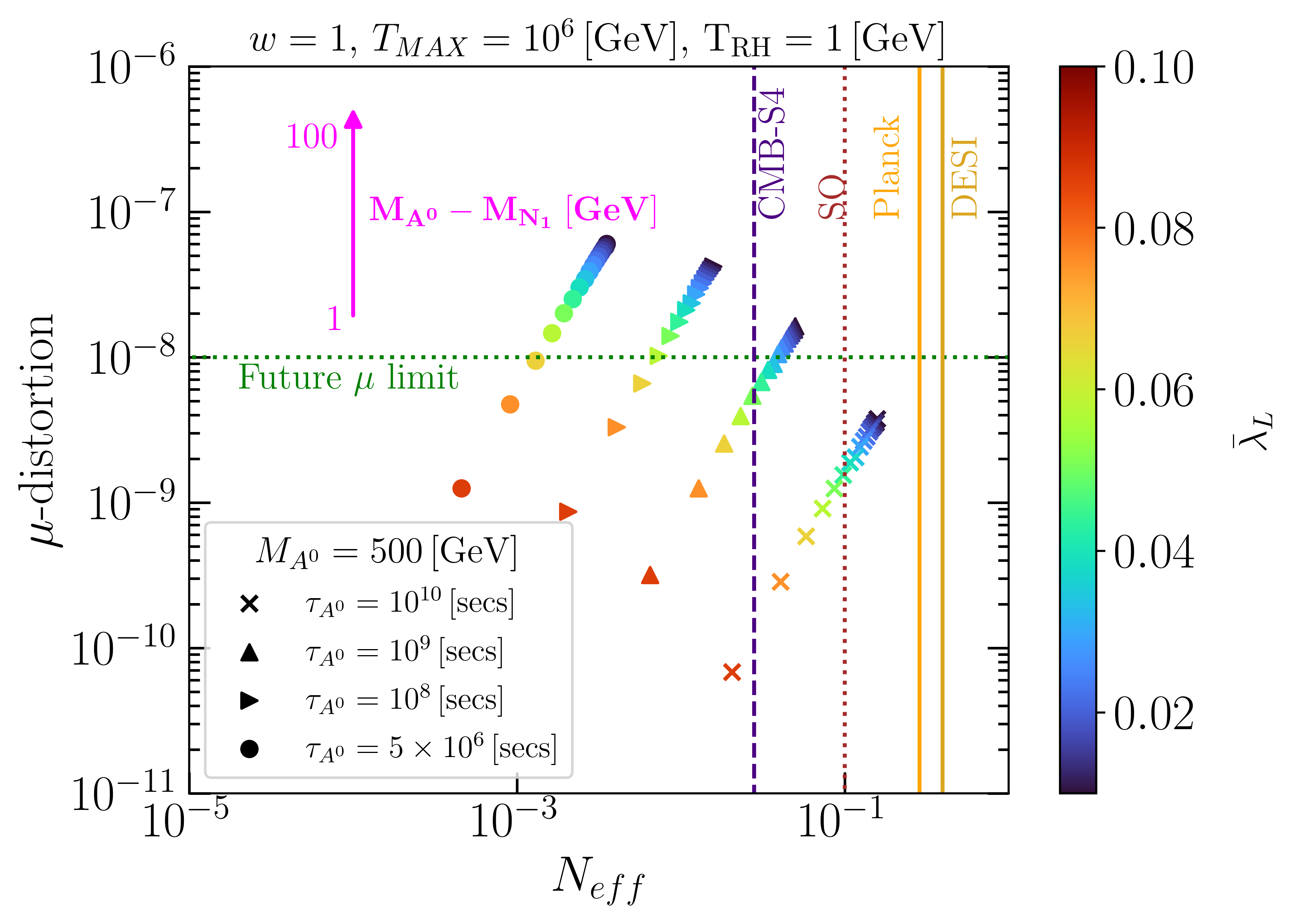}\label{Fig:laml_CMB4b}}
}
\caption{ (a) Correlation between $\bar{\lambda}_L$ and $\Omega_{A^0}h^2$ for different values of $M_{A^0}$, fixing $T_{RH}=1\,\rm GeV$. (b) $\mu-$ distortion vs $N_{eff}$ for $M_{A^0}=500\,\rm GeV$ and  different values of $\tau_{A^0}$. The colorbar indicates the variation of $\bar{\lambda}_L$. We have fixed $M_{H^0}-M_{A^0}=2\,\rm{GeV},\, \text{and}\, M_{H^\pm}-M_{H^0}=10\,\rm{GeV}$. All the points satisfy the relic density limit  within $2\sigma$ and  the best-fit neutrino oscillation data.}
\label{Fig:laml_CMB}
\end{figure}

In Fig.~\ref{Fig:mu_tau3a}, we show the correlation between $T_{RH}$ and $\mu$ distortion for $M_{A^0}$ varying in the range $200-500\,\rm GeV$, fixing $\tau_{A^0}=5\times10^6\,\rm secs$. For a fixed $M_{A^0}$, moving from bottom to top along the corresponding curve, $M_{A^0}-M_{N_1}$ increases from $1$ to $100\,\rm GeV$, as indicated by the magenta arrow. The energy transfer rate in an electron-positron pair, which determines the extent of $\mu$-distortion, scales as $Y_{A^0}$ for co-annihilation between the injected  and background neutrinos and as $Y_{A^0}^2$ for injected neutrino pair annihilation. Thus when the injected neutrino energy exceeds the co-annihilation threshold, the co-annihilation channel dominates over the injected-neutrino pair annihilation.  
This is always the case in the  chosen parameter space since  $E_{\nu}^\star\gtrsim 1 \,\rm GeV$. Consequently, the energy transfer rate is primarily determined by the linear dependence on $Y_{A^0}$. Moreover, the $\mu$-distortion arising from $M_{A^0}=200\,\rm GeV$ is always larger than that for larger values of $M_{A^0}$, since $Y_{A^0}$ decreases as $M_{A^0}$ increases (see Fig.~\ref{Fig:Neff_BP2a}). Interestingly, our results reveal a promising window for probing the parameter space through future $\mu$-distortion experiments: for $M_{A^0}-M_{N_1}\gtrsim\mathcal{O}(40\,\rm GeV)$, values of $\mu\geq10^{-8}$ can be  reached, even though none of the corresponding points can be probed by future $N_{\rm eff}$ measurements, as shown in Fig.~\ref{Fig:Neff_BP2b}. This highlights the complementary sensitivity of CMB spectral distortions and establishes $\mu$-distortion as a potentially powerful probe of parameter space that remains inaccessible to conventional $N_{\rm eff}$ constraints. In Fig.~\ref{Fig:mu_tau3b}, we show the correlation between $\tau_{A^0}$ and $\mu$-distortion 
 for $M_{A^0}=200\,\rm GeV$ and $500\,\rm GeV$, and  $M_{A^0}-M_{N_1}=100\,\rm GeV$ and $50\,\rm GeV$. For $10^6\,\rm secs\leq\tau_{A^0}\leq10^8\,\rm secs$, all the points can be probed by future $\mu$-distortion measurements. A comparison of Fig.~\ref{Fig:mu_tau3b} and Fig. ~\ref{Fig:Neff_BP2b}  further shows that 
 for $M_{A^0}=200\,\rm GeV$ and a mass splitting of $100\,\rm GeV$, the region with $\tau_{A^0}>10^7\,\rm s$ can be probed by both future $N_{\rm eff}$ and $\mu$-distortion measurements. Note however that $\tau_{A^0}>10^9\,\rm s$,  is already excluded by $N_{\rm eff}$ constraints (see Fig.~\ref{Fig:Neff_BP2c}). This is not the case for other parameter choices where the region $\tau_{A^0}>10^9\,\rm s$ lie beyond the reach of future $\mu$-distortion measurements although it remains within the reach of future $N_{\rm eff}$ measurements (see Fig.~\ref{Fig:Neff_BP2c}).

In Fig.~\ref{Fig:laml_CMB4a}, we show the correlation between $\bar{\lambda}_L$ and $\Omega_{A^0}h^2$ for different values of  $M_{A^0}$ while  fixing $T_{RH}=1\,\rm GeV$. We observe that $\Omega_{A^0}h^2$ decreases as $\bar{\lambda}_L$ increases. This is because increasing $\bar{\lambda}_L$ enhances the $A^0$ annihilation rate mediated by the SM Higgs, keeping $A^0$ in thermal equilibrium longer and consequently suppressing its number density after freeze-out. Furthermore, for $M_{A^0}$ in the range $200$–$400\,\rm GeV$, $\Omega_{A^0}h^2$  always lies well below $0.12$, thus making it impossible to account for the  relic density of $N_1$. 
However, for $M_{A^0}=500,\rm GeV$ and $\bar{\lambda}_L$ in the range $10^{-2}$–$10^{-1}$, we find that $\Omega_{A^0}h^2\gtrsim 0.12$ thus allowing to recover the observed DM relic density when the  mass splitting $M_{A^0}-M_{N_1}$ lies in the range $1$–$100\,\rm GeV$. Our results highlight the importance of a low reheating temperature in obtaining the observed DM relic density especially for $M_{A^0}$ in the range $200$–$400\,\rm GeV$.

In Fig.~\ref{Fig:laml_CMB4b}, we show a scatter plot in the $N_{eff}-\mu$ plane for $M_{A^0}=500\,\rm GeV$ and different values of $\tau_{A^0}$. The colorbar indicates the variation of $\bar{\lambda}_L$. As mentioned above,  for small values of $\bar{\lambda}_L$, when  $\tau_{A^0}$ increases from $5\times10^6$ to $10^{10}\,\rm secs$, the parameter space passes through distinct regions of experimental sensitivity. For $\tau_{A^0}= 5\times10^6\,\rm secs$, the corresponding points can be probed by future $\mu$-distortion experiments. Around $\tau_{A^0}= 10^9\,\rm secs$, they become accessible to both future $\mu$-distortion and $N_{\rm eff}$ measurements, whereas at $\tau_{A^0}=10^{10}\,\rm secs$, they can be probed only through $N_{\rm eff}$. Nevertheless, for $\bar{\lambda}_L=0.1$, all the  parameter space remains beyond the reach of both future $\mu$-distortion and $N_{\rm eff}$ experiments.
\begin{table}[]
    \centering
    \footnotesize
    \setlength{\tabcolsep}{4pt}
    \begin{tabular}{lll}
        \toprule
        \textbf{Analysis} & \textbf{Signature Description} & \textbf{Processes} \\
        \midrule
        \texttt{ATLAS-EXOT-2018-06}~\cite{ATLAS:2021kxv} & \texttt{Monojet Searches} & \(pp\to H_iH_j+j_{\rm ISR}\) \\
        \texttt{CMS-SUS-23-003}~\cite{CMS:2025ttk} & \texttt{high-\(p_T\) ISR, multi-leptons (jets)} & \(pp\to H^+H^-\) \\
        \texttt{CMS-EXO-20-004}~\cite{CMS:2021far} & \texttt{Monojet Searches} & \(pp\to H_iH_j+j_{\rm ISR}\) \\
        \texttt{ATLAS-SUSY-2018-16}~\cite{ATLAS:2019lng} & \texttt{2 soft leptons + jets} & \(pp\to H^+H^-,\,H^0A^0\) \\
        \texttt{CMS-SUS-18-004}~\cite{CMS:2021edw} & \texttt{2--3 soft leptons} & \(pp\to H^\pm H^0\) \\
        \texttt{ATLAS-SUSY-2016-07}~\cite{ATLAS:2017mjy} & \texttt{0 leptons + jets} & \(pp\to H_iH_j+4j\) \\
        \texttt{CMS-EXO-19-010}~\cite{CMS:2020atg} & \texttt{disappearing track} & \(pp\to H^+H^-,\,H^\pm H_i; H^\pm\to A^0\pi^\pm\) \\
        \texttt{ATLAS-SUSY-2016-06}~\cite{ATLAS:2017oal} & \texttt{disappearing track} & \(pp\to H^+H^-(H_i); H^\pm\to A^0\pi^\pm\) \\
        \texttt{ATLAS-SUSY-2018-19}~\cite{ATLAS:2022rme} & \texttt{disappearing track} & \(pp\to H^+H^-(H_i); H^\pm\to A^0\pi^\pm\) \\
        \texttt{ATLAS-SUSY-2016-32}~\cite{ATLAS:2019gqq} & \texttt{HSPC} & \(pp\to H^+H^- (H_i);\,\, H^\pm~\text{(long-lived)} \) \\
        \texttt{CMS-EXO-13-006}~\cite{CMS:2015lsu} & \texttt{HSPC} & \(pp\to H^+H^- (H_i);\,\, H^\pm~\text{(long-lived)} \) \\
        \bottomrule
    \end{tabular}
    \caption{Lists of relevant LHC searches and their corresponding production processes used to constrain the parameter points using \texttt{SModelS-3.2.1}~\cite{Altakach:2024jwk}. Here, \(H_i,H_j\in\{H^0,A^0,H^\pm\}\).}
    \label{tab:Search_LHC}
\end{table}

\begin{figure}[]
\centering
\mbox{
\subfigure[]{\includegraphics[width=12cm, height=6cm]{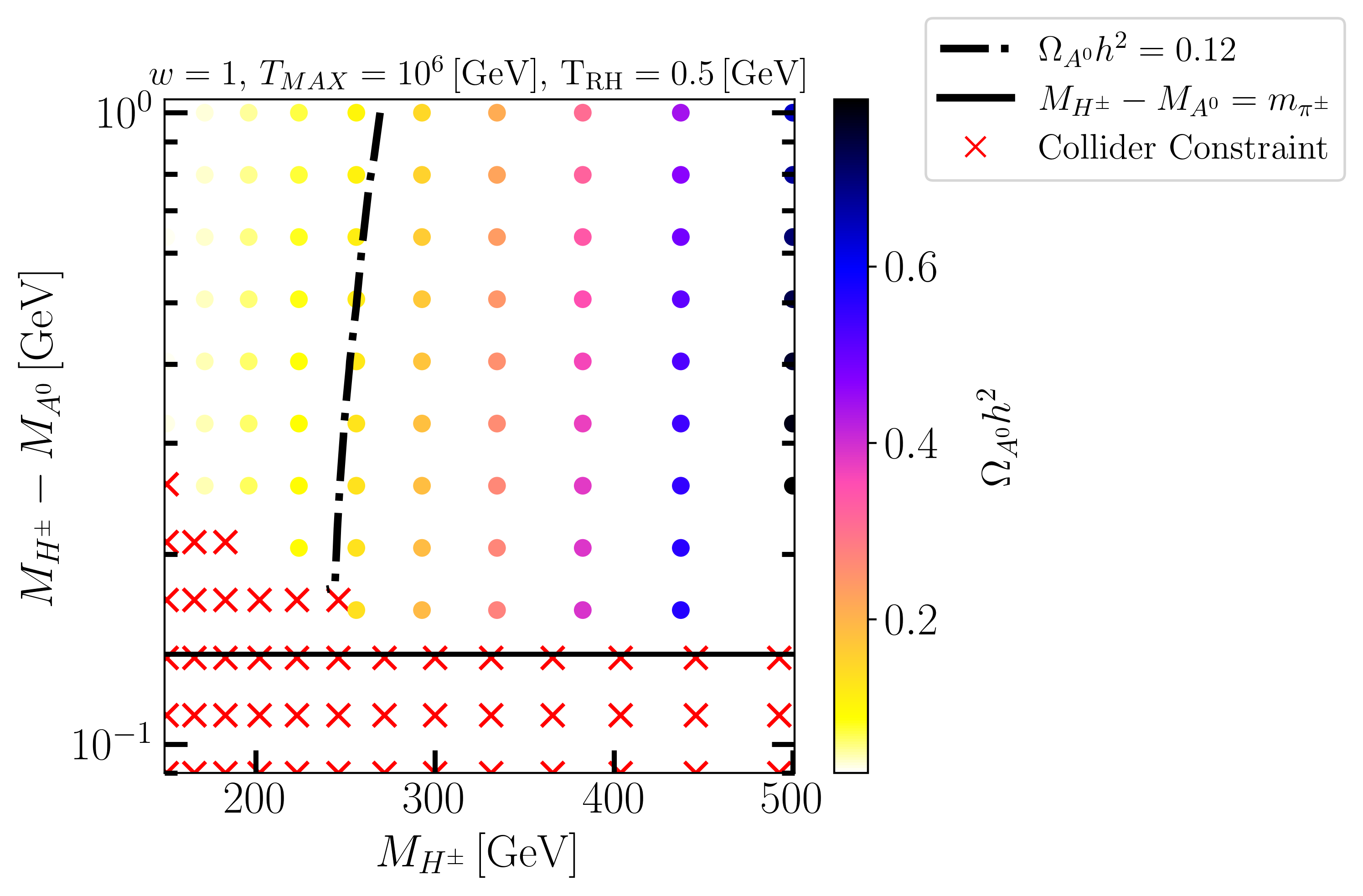}\label{Fig:correlation_plot5a}\label{Fig:correlation_plot5a}
}
}
\mbox{
\subfigure[]{\includegraphics[width=0.5\textwidth]{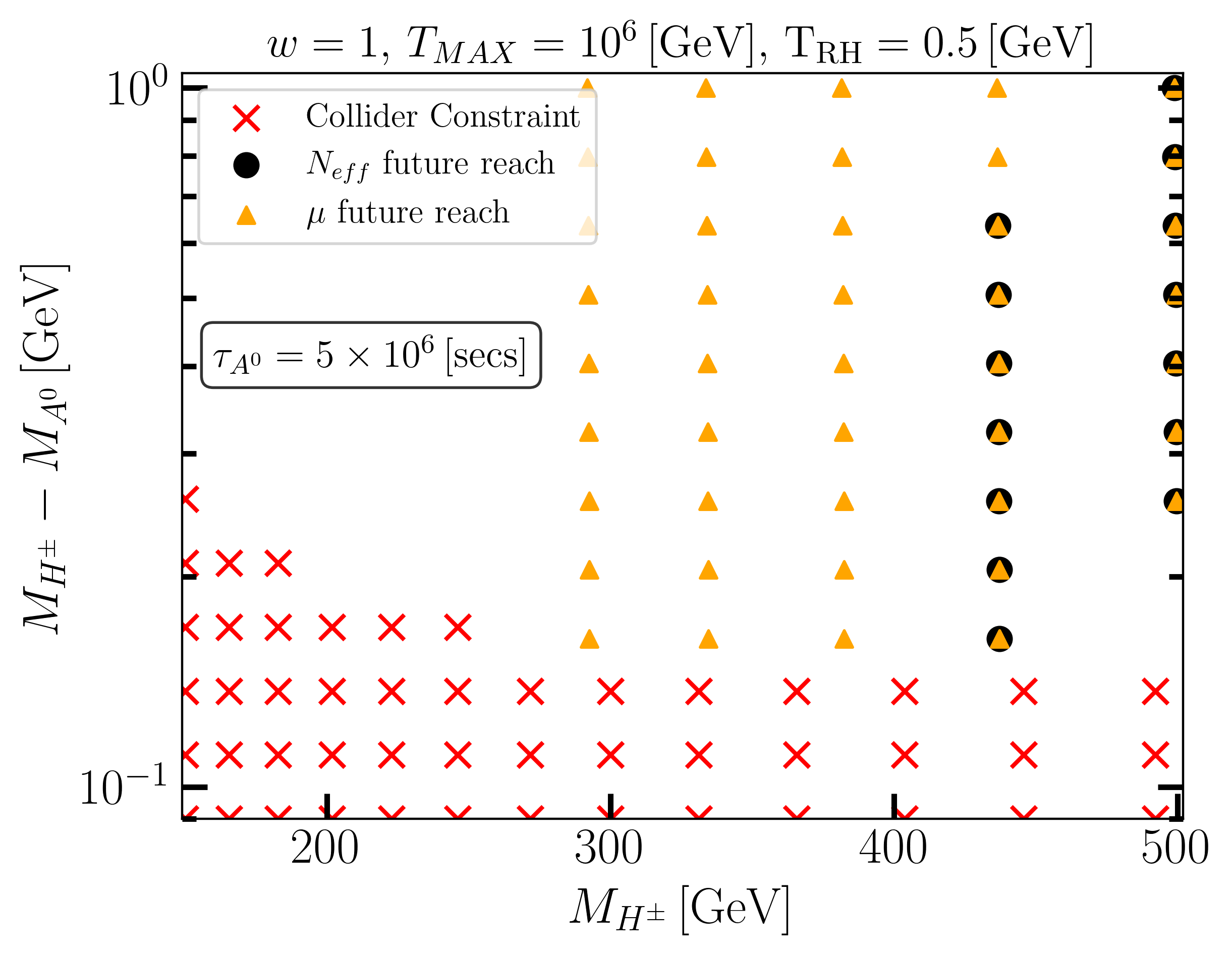}\label{Fig:correlation_plot5b}\label{Fig:correlation_plot5b}
}
\subfigure[]{\includegraphics[width=0.5\textwidth]{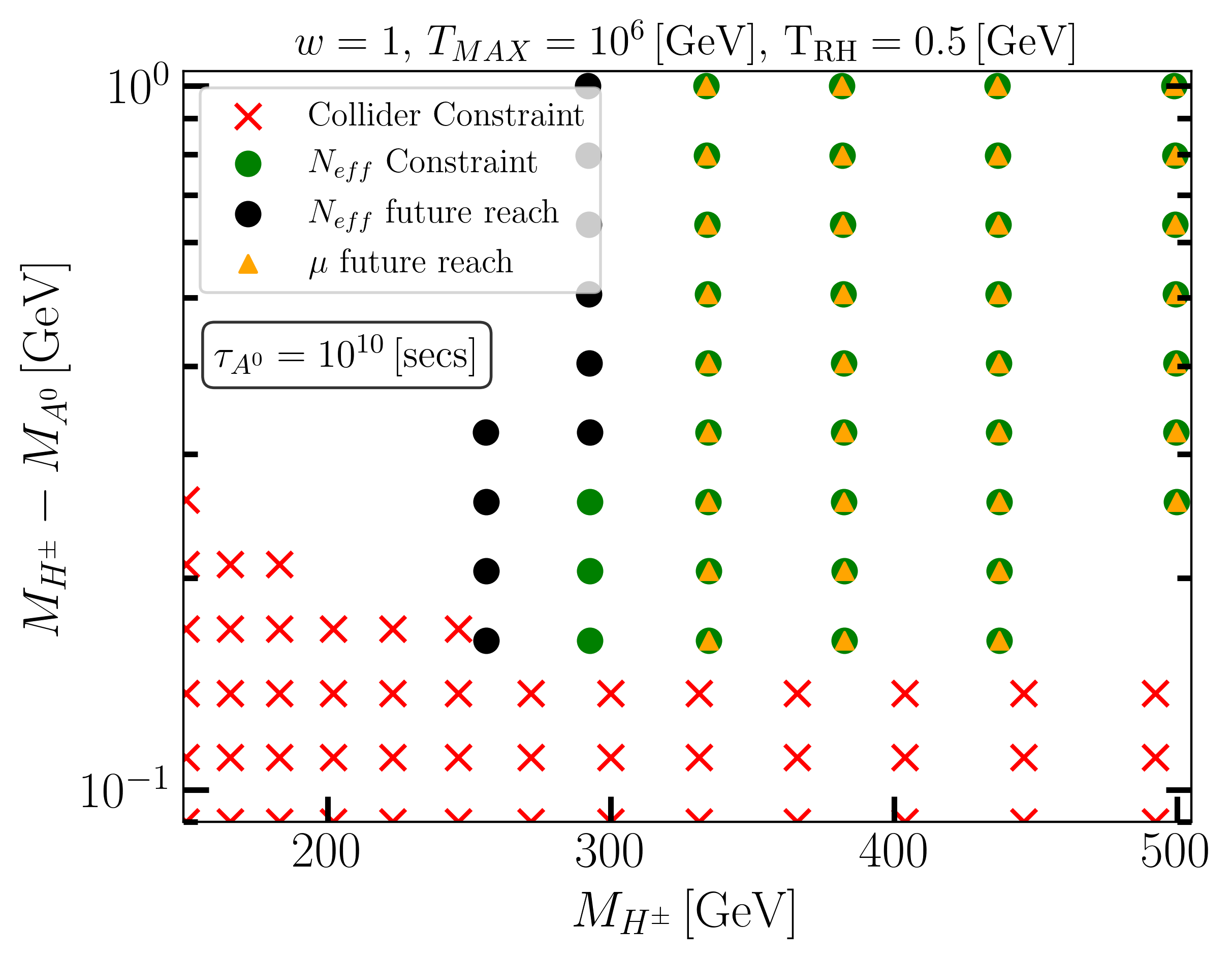}\label{Fig:correlation_plot5c}\label{Fig:correlation_plot5c}
}
}
\caption{(a) Scatter plot in the $M_{H^\pm}-(M_{H^\pm}-M_{A^0})$ plane, assuming $M_{H^\pm}=M_{H^0}$, $T_{RH}=1\,\rm GeV$ and $\bar{\lambda}_L=0$. The color pallet corresponds to $\Omega_{A^0}h^2$ at $T=1\,\rm MeV$. The black dot-dashed and solid lines indicate $\Omega_{A^0}h^2=0.12$ and $M_{H^\pm}-M_{A^0}=m_{\pi^\pm}$. 
(b) Constraints from colliders in the $M_{H^\pm}-(M_{H^\pm}-M_{A^0})$ plane, points that  can be  probed from  future measurements of $N_{\rm eff}$ and $\mu-$distortion measurements are shown  as circles and triangles respectively, here $\tau_{A^0}=5\times 10^6 sec$. (c) Same as (b) with 
$\tau_{A^0}=10^{10} sec$.
}
\label{Fig:correlation_plot}
\end{figure}
\section{Collider Signatures: Heavy Stable Charge Particle and Disappearing Track Signatures}\label{Sec:collider}
We test the viability of our model against a wide range of LHC searches for new scalars using \texttt{SModelS-3.2.1}~\cite{Altakach:2024jwk}, as implemented within \texttt{micrOMEGAs-7.1}~\cite{Belanger:2026asz}. We summarize the relevant searches and their corresponding signatures in Table~\ref{tab:Search_LHC}. Note that the decay of the charged Higgs is dominantly $H^+\rightarrow A^0 W^*$ as in the inert doublet model since the channel $H^+\rightarrow l^+ N_1$ is suppressed by the very small Yukawa $Y^\nu$. We find that none of the searches listed in the table constrains the parameter points discussed in Sections~\ref{sec:DMobs}-\ref{Sec:Como_Sig},  particularly  since we consider a mass-splitting regime of a few GeV's. This  large mass splitting  ensures that $H^\pm$ decays promptly rather than behaving as a long-lived charged particle. Consequently, searches for heavy stable charged particles (HSCPs) and disappearing tracks are not relevant in the parameter region considered earlier.

When the mass splitting between  charged Higgs $H^\pm$ and $A^0$ is $\mathcal{O}(100\,\rm{MeV})$, $H^\pm$ can be long-lived. In our model, long-lived $H^\pm$ can be searched either by searching for a heavy stable charged particle (HSCP) or by searching for a disappearing-track signature in colliders. To compute the lifetime of $H^\pm$, we include effective ${H}^\pm A^0\pi^\pm$ and ${H}^\pm H^0\pi^\pm$ interactions in our \texttt{CalcHEP}~\cite{Belyaev:2012qa} model file, following Refs.~\cite{Belanger:2021lwd,Belyaev:2016lok,Belyaev:2020wok,Chen:1996ap}. These interactions encode the non-perturbative $W$--$\pi$ mixing described by
\begin{equation}
{\cal L}= \frac{g f_\pi}{2\sqrt{2}} W_\mu^+\partial^\mu \pi^-,
\end{equation}
where $f_\pi=130$~MeV. The resulting $H^+A^0\pi^-$ vertex is
\begin{equation}
\frac{g^2 f_\pi}{4\sqrt{2}M_W^2}
(p_{H^+}-p_{A^0})\cdot p_\pi.
\end{equation}
We include the effective vertex for $m_{\pi^\pm}\leq M_{H^\pm}-M_{A^0}\leq500~\mathrm{MeV}$, where $m_{\pi}=139\,\rm{MeV}$ is charged pion mass. In this mass splitting range, we found that $H^\pm\to\pi^\pm A^0$ dominates over the 3-body decay, i.e $H^\pm\to u\bar{d} A^0$, via an off-shell W-boson, as $\Gamma(H^\pm\to\pi^\pm A^0)\gg \Gamma(H^\pm\to u\bar{d} A^0)$~\cite{Chen:1995yu,Chen:1996ap,Chen:1999yf}. The resulting pion is soft and therefore challenging to reconstruct experimentally leading to a disappearing-track signature. In contrast, for $M_{H^\pm}-M_{A^0}<m_{\pi^\pm}$, $H^\pm$ can traverse the entire detector, giving rise to a signature targeted by HSCP searches. Typically, the HSCP searches constrain lifetime of  $H^\pm$  from about $\mathcal{O}(1\,\rm m)$ onwards, and disappearing-tracks from $\mathcal{O}(1\,\rm cm)-\mathcal{O}(1\,\rm m)$~\cite{Cerdeno:2026qio,Alguero:2021dig,Ambrogi:2018ujg}. 

In Fig.~\ref{Fig:correlation_plot5a}, we show a scatter plot in the $M_{H^\pm}-(M_{H^\pm}-M_{A^0})$ plane, assuming $M_{H^\pm}=M_{H^0}$, $T_{RH}=1\,\rm GeV$ and $\bar{\lambda}_L=0$. The color pallet corresponds to $\Omega_{A^0}h^2$ calculated at $T=1\,\rm MeV$. The black dot-dashed and solid line indicates $\Omega_{A^0}h^2=0.12$ and $M_{H^\pm}-M_{A^0}=m_{\pi^\pm}$.
When the mass splitting between $M_{H^\pm}$ and $M_{A^0}$ is smaller than the charged pion mass, the charged scalar becomes long-lived and is constrained by Heavy Stable Charged Particle (HSCP) searches. For mass splittings larger than the pion mass, the charged scalar can decay into a pion, leading to a disappearing-track signature. Red crosses indicate the corresponding collider-excluded points. We computed $\Omega_{A^0}h^2$ for only those points that are not restricted by collider searches. The points left of the black dot-dashed line correspond to $\Omega_{A^0}h^2<0.12$ and cannot produce the observed DM relic density. Therefore, we tested the points to the right of the dot-dashed line against current and future reach on $N_{eff}$ and $\mu$-distortion measurements for $\tau_{A^0}=5\times10^6\,\rm secs$ and $ 10^ {10} \,\rm secs$. In Fig.~\ref{Fig:correlation_plot5b}, for $\tau_{A^0}=5\times10^6\,\rm secs$, we see that future $\mu$-distortion measurements (orange points) can probe all points to the right of the black dot-dashed line, and also that $M_{H^\pm}>400\,\rm GeV$ can be probed in both future $N_{eff}$ (black points) and $\mu$-distortion measurements. In contrast, for $\tau_{A^0}=10^{10}\,\rm secs$, as shown in Fig.~\ref{Fig:correlation_plot5c}, the current measurement of $N_{\rm eff}$ (green points) rules out a large parameter space that interestingly overlaps with future $\mu$-distortion sensitivity.
\section{Summary and Discussion}\label{sec:conclusion}

In this work, we have investigated the cosmological signatures of dark matter scenarios featuring a late-decaying particle, focusing in particular on its impact on the effective number of relativistic species, \(N_{\rm eff}\), and on the chemical potential distortion, \(\mu\), 
of the Cosmic Microwave Background (CMB) spectrum. We have shown that the late decay of an unstable particle into dark matter and neutrinos can lead to a non-negligible injection of energy into the neutrino sector. 
Our analysis was performed within the framework of the scotogenic model, assuming a non-standard cosmological history characterised by a kination-dominated era. This modification of the early-Universe expansion history allows the observed dark matter relic abundance to be reproduced in regions of the model parameter space where the standard cosmological scenario would lead to underabundant dark matter.  Although our analysis is carried out within this specific framework, the main conclusions can be extended to other dark matter scenarios featuring similar late-decaying particles.
Our results show that depending on the mass and lifetime of the unstable particle and on the mass of dark matter, future CMB experiments could  detect a signal in \(N_{\rm eff}\), in \(\mu\)-distortions, or potentially in both 
observables simultaneously. These complementary signatures provide a way to probe the properties of dark matter and the thermal history of the early Universe. 

In our numerical analysis, we have focused on a mass range of \(200\)--\(500\) GeV for the lightest scalar component of the inert doublet. This range was chosen to optimise  the cosmological signals 
while offering good prospects for collider searches.  For larger scalar masses, we generally expect the predicted cosmological signals to become suppressed, whereas lighter scalar masses could lead to enhanced signals 
due to their larger production yields. However their larger production yields can also lead to more substantial energy injection during the early Universe, potentially bringing them into conflict with existing BBN bounds. 
We have also discussed the possibility of probing part of the parameter space through collider searches at the LHC.

Overall, our study demonstrates that future CMB experiments could provide a unique opportunity to probe aspects of dark matter production and the early-Universe thermal history that are difficult to access through 
 conventional particle physics searches alone.

\section*{Acknowledgements}
The work of AR is supported by Basic Science Research Program through the National Research Foundation of Korea (NRF) funded by the Ministry of Education through the Center for Quantum Spacetime (CQUeST) of Sogang University (RS-2020-NR049598) and by the Ministry of Science and ICT with grant number RS-2025-24523022.

\appendix
\section{Analytic derivation of collision term, injected neutrino distribution and energy transfer rate} \label{appendix}

\subsection{Collision term  for $A^0 \to N_1 \nu$}\label{Appendix:Collision_Term_Calc}
We first consider the contribution to the neutrino phase-space distribution from the decay process $A^0\rightarrow N_1+\nu$, characterized by the decay width $\Gamma_{A^0}$, which contributes as source term to the neutrino phase space distribution. The collision term can be written as,
\begin{equation}
    \mathcal{C}_{\nu}[f^{\rm inj}_\nu]
=\frac{1}{2E_{\nu}}\int \left(\prod_{i=A^0,N_1} \frac{d^3p_i}{(2\pi)^32E_i} \right) (2\pi)^{4} \delta^{4}(p_{A^0}-p_{N_1}-p_{\nu}) |\mathcal{M}|^{2} f_{A^0} 
\, .
\end{equation}
After integrating over the momentum of $N_1$, we obtain,
\begin{equation}
    \mathcal{C}_{\nu}[f^{\rm inj}_\nu]=\frac{\pi|\mathcal{M}|^{2}}{2E_{N_1}E_{\nu}}\int \frac{d^3p_{A^0}}{(2\pi)^32E_{A^0}} \delta(E_{A^0}-E_N-E_\nu) f_{A^0}
\end{equation}
Since $A^0$ is non-relativistic during its decay, we take,
\begin{align}
    E_{A^0}&\simeq M_{A^0},\\
     f_{A^0}&=(2\pi)^3\delta^3(p_{A^0})n_{A^0}.
\end{align}
the collision term becomes,
\begin{equation}
    \mathcal{C}_{\nu}[f^{\rm inj}_\nu]=\frac{\pi|\mathcal{M}|^{2}n_{A^0}}{4E_{N_1}E_{\nu}E_{A^0}}\delta(E_{A^0}-E_N-E_\nu) 
\end{equation}
For a two-body decay in the rest frame of $A^0$, the neutrino is produced with a fixed energy and is given by,
\begin{equation}
    E^\star_\nu=\frac{M^2_{A^0}-M^2_{N_1}}{2M_{A^0}}
\end{equation}
Therefore,
\begin{equation}
    \delta(M_{A^0}-E_N-E_\nu)=\frac{E_{N_1}}{M_{A^0}}\delta(E^\star_\nu-E_\nu),
\end{equation}
and hence
\begin{equation}
    \mathcal{C}_{\nu}[f^{\rm inj}_\nu]=\frac{\pi|\mathcal{M}|^{2}n_{A^0}}{4E_{\nu}M^2_{A^0}}\delta(E^\star_\nu-E_\nu)
\end{equation}

The parent number density evolves due to both the expansion of the Universe and the decay of $A^0$. Defining the yield of $Y_{A^0}$ at neutrino decoupling as
\begin{equation}
    Y_{A^0}\equiv\frac{n_{A^0}(t_{\rm dec})}{s(t_{\rm dec})},
\end{equation}
the number density at a later time is
\begin{equation}
    n_{A^0}(t)
    =
    n_{A^0}(t_{\rm dec})
    \left(\frac{a_{\rm dec}}{a}\right)^3
    e^{-\Gamma_{A^0}(t-t_{\rm dec})}.
\end{equation}
Here, the factor $(a_{\rm dec}/a)^3$ accounts for the dilution of the number density due to cosmic expansion, while the exponential factor describes the decay of the unstable parent. Since entropy is conserved after neutrino decoupling, $s\propto a^{-3}\propto T^3$, and hence
\begin{equation}
    n_{A^0}(t_{\rm dec})=Y_{A^0}c_sT_{\nu{\rm dec}}^3,
    \qquad
    \left(\frac{a_{\rm dec}}{a}\right)^3
    =
    \left(\frac{T}{T_{\nu{\rm dec}}}\right)^3.
\end{equation}
Therefore,
\begin{equation}
    \boxed{
    n_{A^0}(t)
    =
    Y_{A^0}c_sT^3
    e^{-\Gamma_{A^0}(t-t_{\rm dec})}.
    }
\end{equation}
Introducing,
\begin{align}
x\equiv \frac{T_{\nu\rm dec}}{T}
\end{align}
and assuming radiation domination with approximately constant relativistic degrees of freedom, we have
\begin{equation}
    \frac{dx}{dt}=Hx=\frac{T^2_{\nu\rm dec}}{xM^\star_p}
\end{equation}
which gives
\begin{equation}
    t-t_{dec}=\frac{M^\star_p}{2T^2_{\nu\rm dec}}(x^2-1)
\end{equation}
Writing the initial abundance as
\begin{equation}
    n_{A^0}(t)=Y_{A^0}c_sT^3e^{\frac{\Gamma_{A^0}M^\star_p}{2T^2_{\nu\rm dec}}(1-x^2)}=Y_{A^0}c_sT^3e^{\eta(1-x^2)},
\end{equation}
where,
\begin{equation}
    \boxed{
    \eta=\frac{\Gamma_{A^0}M^\star_p}{2T^2_{\nu\rm dec}}
    }
\end{equation}
Substituting this into the collision term gives,
\begin{align}
    \mathcal{C}_{\nu}[f^{\rm inj}_\nu]=\frac{\pi|\mathcal{M}|^{2}}{4E_{\nu}M^2_{A^0}}Y_{A^0}c_sT^3e^{\eta(1-x^2)}\delta(E^\star_\nu-E_\nu) 
\end{align}
Finally, using the relation between the squared matrix element and the decay width,
\begin{equation}
    |\mathcal{M}|^{2}=\frac{16\pi M^3_{A^0}}{M^2_{A^0}-M^2_{N_1}}\Gamma_{A^0}
\end{equation}
together with,
\begin{equation}
    M^2_{A^0}-M^2_{N_1}=2M_{A^0}E^\star_\nu
\end{equation}
we obtain,
\begin{align}
    \boxed{
    \mathcal{C}_{\nu}[f^{\rm inj}_\nu]=\frac{2\pi^2\Gamma_{A^0}Y_{A^0}c_sT^3}{E^{\star}_{\nu}E_{\nu}}e^{\eta(1-x^2)}\delta(E^\star_\nu-E_\nu)
    }
\end{align}

\subsection{Injected neutrino distribution $f_{\nu}$}
\label{Appendix:Dis_Func_Calc}
The Boltzmann equation for the evolution of injected neutrino distribution function can be written as,
\begin{align}
\frac{d f_{\nu}^{\rm inj}}{d  t} =\frac{\partial f_{\nu}^{\rm inj}}{\partial t}-H|\p|\frac{\partial f_{\nu}^{\rm inj}}{\partial |\p|}=\mathcal{C}_\nu[f^{\rm inj}_\nu] \,.
\end{align}
By imposing the initial condition $f_{\nu}^{\rm inj}(t_{\nu,\rm dec},|\p|)=0$, the solution is obtained by integrating the collision term along the phase-space characteristic:
\begin{align}
f_{\nu}^{\rm inj}(t, |\p|) &= \int_{t_{\nu,\rm dec}}^{t} dt' \, \mathcal{C}_{\nu} (t',|\p|(t')) \nonumber\\
&=\int_{t_{\nu, \rm dec}}^{t} dt'\left[ \frac{2\pi^2\Gamma_{A^0}Y_{A^0}c_sT^3}{E^{\star}_{\nu}E_{\nu}}e^{\eta(1-x^2)}\delta(E^\star_\nu-E_\nu) \right]
\label{Appen_Eqn:Collision_term}
\end{align}
By defining the dimensionless variables, 
\begin{align}
 r=\frac{|{\bf E}_\nu|}{T}\,,  \quad r_{\star}= \frac{E^{\star}_\nu}{T_{\nu\rm dec}}
\end{align}
We can rewrite the Eqn.~\ref{Appen_Eqn:Collision_term} as,
\begin{align}
    f_{\nu}^{\rm inj}(r, x) &=\frac{2\pi^2Y_{A^0}\Gamma_{A^0}M^{\star}_pc_s}{T^{2}_{\nu,\rm dec}rr_{\star}}\int_{1}^{x}dx^\prime e^{\eta(1-x^{\prime2})}\delta(r-r_\star x^\prime)\,\nonumber\\
    &=\frac{4\pi^2 \eta Y_{A^0}c_s}{rr^2_{\star}}e^{\eta\left(1-\frac{r^2}{r^2_\star}\right)}\theta(r-r_\star)\theta(xr_\star-r) \quad (\rm as\, \Gamma_{A^0}M^\star_p=2\eta T^2_{\nu,\rm dec})
\end{align}
Hence, $f_{\nu}^{\rm inj}$ can be written as
\begin{align}
    \boxed{
    f_{\nu}^{\rm inj}(r, x)=\frac{4\pi^2 \eta Y_{A^0}c_s}{rr^2_{\star}}e^{\eta\left(1-\frac{r^2}{r^2_\star}\right)}\Theta(r,r_\star,x).
    }
\end{align}

\subsection{Energy  transfer rate to electron-positron pair}
\label{Appendix:Energy_Tranfer_rate}
 The energy transfer rate into the electron-positron pair is then given by 
\begin{align}
	\frac{{\rm d}\rho_{e\bar e}^{\rm inj(bg)}}{{\rm d}t}&=\int \left(\prod_{i=\nu,\bar \nu, e,\bar e}\frac{{\rm d}^3 {\bf p}_i}{(2\pi)^3 2E_i} \right) \delta^4(p_\nu+p_{\bar \nu_{}}- p_e-p_{\bar e})|\mathcal{M}|^2_{\nu\to e} (E_e+E_{\bar e})f^{\rm inj}_\nu f^{\rm inj(bg) }_{\bar \nu}\nonumber\\
    &=\frac{G_{\rm F}^2}{3\pi}\int \left(\prod_{i=\nu,\bar \nu}\frac{{\rm d}^3 {\bf p}_i}{(2\pi)^3 2E_i} \right)(E_\nu+E_{\bar \nu})f^{\rm inj}_\nu f^{\rm inj(bg) }_{\bar \nu}\left(\left(s^2+2sm_e^2\right)g_W-3sm_e^2\right)\theta(s-4m_e^2),
    \label{Eqn:Appendix_rate2}
\end{align}
where the total energy transfer is  given by $\rho_{e\bar e}=\rho_{e\bar e}^{\rm inj}+\rho_{e\bar e}^{\rm bg}$, $G_{\rm F}=1.167\times 10^{-5}~\text{GeV}^{-2}$ is the Fermi constant, while $s=2E_{\nu}E_{\bar\nu}(1-\cos\theta)$ is the Mandelstam variable corresponding to the center-of-mass energy squared of the neutrino-antineutrino system. The factor $\theta(s-4m_e^2)$ ensures that the process is kinematically allowed only when the center-of-mass energy exceeds the electron-positron production threshold. We define $g_W\equiv \left(1+4\sin\theta_W^2+8\sin\theta_W^4\right)$, where $\theta_W$ is the weak mixing angle and $\sin\theta_W^2\approx 0.23$. Note that in Eqn.~\ref{Eqn:Appendix_rate2}, we have used the electron-
positron phase-space integration in the center-of-mass frame, which is given by,
\begin{align}
	\int {\rm d}\Pi_e {\rm d}\Pi_{\bar e} \,\tilde \delta^4 \left(p\right)|\mathcal{M}|^2_{\nu\to e}
=\frac{G_{\rm F}^2}{3\pi}\left( \left(s^2+2sm_e^2\right)g_W-3sm_e^2\right)\theta(s-4m_e^2).
\end{align}
We can further simplify Eqn.~\ref{Eqn:Appendix_rate2} which can be written as
\begin{align}
    \frac{{\rm d}\rho_{e\bar e}^{\rm inj(bg)}}{{\rm d}t}&=\frac{G_{\rm F}^2}{96\pi^5}\int dE_\nu dE_{\bar \nu}E_{\nu}E_{\bar\nu}(E_\nu+E_{\bar \nu})f^{\rm inj}_\nu f^{\rm inj(bg) }_{\bar \nu}\int_{-1}^{+1}d\cos\theta \left(\left(s^2+2sm_e^2\right)g_W-3sm_e^2\right)\theta(s-4m_e^2)\nonumber\\
    &=\frac{G_{\rm F}^2}{192\pi^5}\int dE_\nu dE_{\bar \nu}(E_\nu+E_{\bar \nu})f^{\rm inj}_\nu f^{\rm inj(bg) }_{\bar \nu}\int_{4m^2_e}^{4E_{\nu}E_{\bar\nu}}ds\left(\left(s^2+2sm_e^2\right)g_W-3sm_e^2\right)\nonumber\\
    &=\frac{G_{\rm F}^2}{192\pi^5}\int dE_\nu dE_{\bar \nu}(E_\nu+E_{\bar \nu})f^{\rm inj}_\nu f^{\rm inj(bg) }_{\bar \nu}\mathcal{K}(E_\nu,E_{\bar\nu}),
\end{align}
where $\mathcal{K}(E_\nu,E_{\bar\nu})$ is given by 
\begin{align}
    \mathcal{K}(E_\nu,E_{\bar\nu})=\frac{64}{3}g_W\left(E^3_\nu E^3_{\bar\nu}-m^6_e \right)+8(2g_W-3)m^2_e\left(E^2_\nu E^2_{\bar\nu}-m^4_e\right)
\end{align}

\bibliographystyle{JHEP}
\bibliography{bibitem.bib}
\end{document}